\RequirePackage{fix-cm}
\documentclass[natbib,smallextended]{svjour3}       
\smartqed  
\usepackage{color}
\usepackage{graphicx}
\usepackage{amssymb,amsmath}
\usepackage{hyperref}
\hypersetup{
    colorlinks,%
    citecolor=blue,%
    linkcolor=blue,%
    urlcolor=blue
}
\usepackage[normalem]{ulem}
\usepackage{aps-bibstyle}  

\newcommand{\ratechap}{Rates Chapter}
\newcommand{\disrupchap}{Disruption Chapter}
\newcommand{\diskchap}{Accretion Disc Chapter}
\newcommand{\simchap}{Simulation Methods Chapter}
\newcommand{\emischap}{Emission Mechanisms Chapter}

\newcommand{\binchap}{Binaries Chapter}

\newcommand{\fmodchap}{Future Modeling Chapter}

\newcommand{\mh}{M_{\rm h}}
\newcommand{\mdotfb}{\dot{M}_{\rm fb}}
\newcommand{\rint}{R_{\rm int}}
\newcommand{\rp}{R_{\rm p}}
\newcommand{\tvisc}{t_{\rm visc}}
\newcommand{\rt}{R_{\rm t}}
\newcommand{\rg}{R_{\rm g}}
\newcommand{\tmin}{t_{\rm min}}
\newcommand{\tcirc}{t_{\rm circ}}
\newcommand{\amin}{a_{\rm min}}
\newcommand{\emin}{e_{\rm min}}

\newcommand{\ecirc}{\epsilon_{\rm{circ}}}

\newcommand{\mstar}{M_{\star}}
\newcommand{\rstar}{R_{\star}}
\newcommand{\ergpers}{\, \rm erg \, \rm s^{-1}}
\newcommand{\erg}{\, \rm erg}
\newcommand{\days}{\, \rm d}
\newcommand{\hours}{\, \rm h}

\newcommand{\K}{\, \rm K}
\newcommand{\gpercc}{\, \rm g \, \rm cm^{-3}}
\newcommand{\csperg}{\, \rm cm^2 \, \rm g^{-1}}
\newcommand{\msunperyr}{\, \rm M_{\odot} \, \rm yr^{-1}}

\usepackage[flushleft]{threeparttable}

\def\msun{\, \mathrm{M}_{\hbox{$\odot$}}}
\def\rsun{\, \mathrm{R}_{\hbox{$\odot$}}}
\def\au{\, \rm{AU}}

 \journalname{ISSI Book on TDEs}

\usepackage{ref_macros} 

\defcitealias{ramirez-ruiz2009}{Ramirez-Ruiz et al. (2009)}
\defcitealias{guillochon2015}{Guillochon et al. (2015)}

\begin{document}

\title{Formation of an Accretion Flow}


\author{C. Bonnerot \and  N.C. Stone}


\institute{C. Bonnerot \at
TAPIR, Mailcode 350-17\\
California Institute of Technology \\
Pasadena, CA 91125 (USA)\\
\email{bonnerot@tapir.caltech.edu}
\and
N.C. Stone \at
Racah Institute of Physics \\
The Hebrew University\\
Jerusalem, 91904 (Israel) \\
\email{nicholas.stone@mail.huji.ac.il}   
}

\date{Received: date / Accepted: date}

\maketitle

\begin{abstract}
 After a star has been tidally disrupted by a black hole, the debris forms an elongated stream. We start by studying the evolution of this gas before its bound part returns to the original stellar pericenter. While the axial motion is entirely ballistic, the transverse directions of the stream are usually thinner due to the confining effects of self-gravity. This basic picture may also be influenced by additional physical effects such as clump formation, hydrogen recombination, magnetic fields and the interaction with the ambient medium. We then examine the fate of this stream when it comes back to the vicinity of the black hole to form an accretion flow. Despite recent progress, the hydrodynamics of this phase remains uncertain due to computational limitations that have so far prevented us from performing a fully self-consistent simulation. Most of the initial energy dissipation appears to be provided by a self-crossing shock that results from an intersection of the stream with itself. The debris evolution during this collision depends on relativistic apsidal precession, expansion of the stream from pericenter, and nodal precession induced by the black hole spin. Although the combined influence of these effects is not fully understood, current works suggest that this interaction is typically too weak to significantly circularize the trajectories, with its main consequence being an expansion of the shocked gas. Global simulations of disc formation performed for simplified initial conditions find that the debris experiences additional collisions that cause its orbits to become more circular until eventually settling into a thick and extended structure. These works suggest that this process completes faster for more relativistic encounters due to the stronger shocks involved. It is instead significantly delayed if weaker shocks take place, allowing the gas to retain large eccentricities during multiple orbits. Radiation produced as the matter gets heated by circularizing shocks may leave the system through photon diffusion and participate in the emerging luminosity. This current picture of accretion flow formation results from recent theoretical works synthesizing the interplay between different aspects of physics. In comparison, early analytical works correctly identified the essential processes involved in disc formation, but had difficulty developing analytic frameworks that accurately combined non-linear hydrodynamical processes with the underlying relativistic dynamics. However, important aspects still remain to be understood at the time of writing, due to numerical challenges and the complexity of this process.

\end{abstract}

\section{Introduction}

In this chapter, we focus on the evolution of the debris around the black hole following stellar disruption, and most importantly the formation of an accretion flow from this matter. These two stages are described below in largely independent sections so that the reader is able to go through them in any order. While our understanding of isolated stream evolution is relatively robust, the later stage of accretion disc formation remains debated due to its greater complexity and associated numerical challenges. For the latter, we therefore present the major advances made so far while emphasizing the uncertainties of these current works.

Following the disruption of the star on its original parabolic trajectory, the debris evolves into an elongated stream due to the spread in orbital energy imparted to the stellar matter by the encounter with the black hole. While half of this gas is unbound and escapes the gravitational attraction of the compact object on hyperbolic orbits, the rest becomes bound and comes back to the stellar pericenter on highly eccentric trajectories. The hydrodynamics at play during this revolution of the stellar debris around the black hole is well understood thanks to the early work of \cite{kochanek1994}, more recently revised by \cite{coughlin2016-structure}. During this phase, the stream gets stretched along its longitudinal direction due to the ballistic in-plane motion of the gas elements that evolve like test particles on a wide range of different elliptical orbits. In the other two transverse directions, the gas motion is usually specified by self-gravity that imposes a thin width such that the stream has a locally cylindrical geometry. The influence of various additional physical ingredients on this basic picture has been explored, including stream fragmentation into individual self-gravitating clumps, the recombination of hydrogen as the stream cools, the magnetic field inherited from the star and the interaction of the debris with the ambient gaseous medium. It is nevertheless fair to say that the basic picture of stream evolution described above is usually not strongly affected by this additional physics.

After the bound part of the stream has completed an entire orbit, it comes back near the black hole where it can start forming an accretion flow. As already mentioned, our understanding of this phase of evolution is less secure than the earlier one due to the more complicated mechanisms involved and the numerical challenges of studying them. The returning stream is strongly compressed at pericenter, causing a nozzle shock whose main effect is to make the gas expand. Initial dissipation is provided by a collision of the stream with itself. The parameters of this collision are specified by the combination of relativistic apsidal precession, expansion from the nozzle shock, and nodal precession due to black hole spin. Local simulations of this interaction \citep[e.g.][]{jiang2016,lu2020} find that the ensuing self-crossing shock dissipates part of the debris kinetic energy and results in an expansion of the gas distribution around the intersection point. However, performing a global study of the disc formation process is very computationally expensive and current investigations therefore rely on simplified initial conditions to circumvent this numerical burden. Most often, this involves either decreasing the stellar eccentricity such that the debris becomes artificially more bound, or reducing the black hole mass to that of an intermediate mass black hole. These numerical works \citep[e.g.][]{hayasaki2013,bonnerot2016-circ,shiokawa2015,bonnerot2020-realistic} find that the shocked gas experiences additional dissipation that causes the debris to move to more circular orbits and to eventually settle into a thick and extended accretion disc. Importantly, this final state may take a long time to be reached, and the gas likely retains moderate to large eccentricities even after this process is completed. The current understanding of this process presents significant improvements compared to early analytic works such as that by \cite{rees1988}, and it significantly differs from the simple picture involving the fast formation of a compact and axisymmetric disc, often assumed in the literature. While the above qualitative description of disc formation is relatively robust, there is so far no clear consensus on the detailed hydrodynamics at play for astrophysically realistic initial conditions. Nevertheless, our understanding of this process has now reached a level sufficient to better identify the main sources of uncertainty and start developing strategies to overcome them.

This chapter is arranged as follows. We start in Section \ref{sec:stream-evolution} by describing the evolution of the stream around the black hole before it comes back to pericenter. The basic gas dynamics is first presented followed by the impact of additional physical processes. In Section \ref{sec:disc-formation}, we explain the current understanding of how the returning debris forms an accretion flow. We start by describing the initial sources of dissipation in a largely analytical fashion and then present the various numerical investigations of the disc formation process along with its main consequences.  Finally, Section \ref{sec:conclusion} contains a summary and our concluding remarks.

\section{Stream evolution prior to pericenter return}
\label{sec:stream-evolution}

\subsection{Basic stream trajectory and geometry}
\label{sec:basic}

The trajectory of the debris stream mostly results from the disruption process discussed in the \disrupchap, but we start by recalling the most important properties below. The star is disrupted by the tidal force of the black hole if its pericenter distance $\rp$ is lower than the tidal radius
\begin{equation}
\rt = \rstar \left(\frac{\mh}{\mstar}\right)^{1/3} = 0.47 \au \left( \frac{\mh}{10^6 \msun} \right)^{1/3},
\label{eq:tidal-radius}
\end{equation}
with the depth this encounter being characterized by the penetration factor $\beta = \rt/\rp$. Here, $\mh$ denotes the mass of the compact object while $\mstar$ and $\rstar$ are the stellar mass and radius, respectively. Note that the tidal radius can change by a factor of a few compared to equation \eqref{eq:tidal-radius} depending on the structure of the disrupted star. As in the above estimate, the numerical values in the remaining of the chapter are usually given assuming a solar-type star with $\mstar = \msun$ and $\rstar = \rsun$ unless stated otherwise. The disruption imparts a spread in specific orbital energy \citep{rees1988,stone2013}
\begin{equation}
\Delta \varepsilon = \frac{G \mh \rstar}{\rt^2},
\label{eq:energy-spread}
\end{equation}
to the debris that makes half of it unbound while the rest gets bound to the black hole. The spread in angular momentum is however negligible \citetext{see for example the figure 2 from \citealt{cheng2014}}, which implies in particular that all the stream elements keep a pericenter distance equal to that of the original star. As the gas keeps orbiting the black hole, it is usually a good approximation to consider its trajectory along the orbital plane as perfectly ballistic. This implies that the gas evolves into an elongated stream with the unbound part escaping to large distances while the bound gas returns to the stellar pericenter after a finite time. In between, a fraction of the debris is marginally-bound, remaining on the parabolic trajectory that the star followed until its disruption. The gas located at the most bound extremity of the stream has a semi-major axis 
\begin{equation}
\amin =  \frac{\rstar}{2} \left(\frac{\mh}{\mstar} \right)^{2/3} = 23 \au \left( \frac{\mh}{10^6 \msun} \right)^{2/3},
\label{eq:semi-major-axis}
\end{equation}
making use of equation \eqref{eq:energy-spread}. It is the first to come back to the black hole following the disruption after a time equal to its orbital period given by
\begin{equation}
\tmin = 2^{-1/2} \, \pi \, \left(\frac{G \mstar}{\rstar^3} \right)^{-1/2} \left(\frac{\mh}{\mstar} \right)^{1/2} = 41 \days \left( \frac{\mh}{10^6 \msun} \right)^{1/2},
\label{eq:fallback-time}
\end{equation}
according to Kepler's third law and equation \eqref{eq:semi-major-axis}. This stream element is highly-eccentric with an eccentricity $\emin = 1- \rp/\amin$ given by
\begin{equation}
1- \emin = \frac{2}{\beta} \left(\frac{\mh}{\mstar}\right)^{-1/3} = 0.02 \, \beta \,\left(\frac{\mh}{10^6 \msun}\right)^{-1/3}
\label{eq:eccentricity}
\end{equation}
using equations \eqref{eq:tidal-radius} and \eqref{eq:semi-major-axis}. The mass fallback rate $\mdotfb$ at which the stellar matter comes back to the black hole is specified by how orbital energy is distributed among the stream. Assuming a flat distribution yields $\mdotfb = \dot{M}_{\rm p} ( t/\tmin )^{-5/3}$ with a peak value of
\begin{equation}
\dot{M}_{\rm p} = \frac{\mstar}{3 \, \tmin} = 3 \msunperyr \left( \frac{\mh}{10^6 \msun} \right)^{-1/2},
\label{eq:fallback-rate}
\end{equation}
such that the integrated amount of returning mass is that $\mstar/2$ of the bound debris. 

The geometry of the stream is well approximated by a cylinder whose thickness and elongation evolve in time. The latter is entirely determined by the ballistic motion of the gas along the orbital plane. In particular, one can specify the dependence of the length $l(R)$ of a stream element of fixed mass on its distance $R$ from the black hole. At early times $t\lesssim \tmin$ when most of the gas still follows a near-parabolic orbit, this length scales as \citep{coughlin2016-structure}
\begin{equation}
l \propto R^2.
\label{eq:length}
\end{equation}
This result was derived by solving the equations of gas dynamics for the marginally-bound part of the stream but it can also be recovered in the following more intuitive way. Immediately after the disruption, the stream is still close to the tidal radius with a size similar to that of the original star, such that $l(\rt) \approx \rstar$. At $t\approx \tmin$, the bulk of the stream reaches a distance of order $\amin$ with most of its elongation being caused by the most bound debris that has returned to the black hole.\footnote{Despite its fast outward motion, the unbound extremity of the stream remains close to the semi-major axis of the most bound gas at $t\approx \tmin$ since its velocity near this location can be approximated by $v_{\rm unb} \approx \sqrt{G \mh / \amin}$ such that the distance reached is $v_{\rm unb} t \approx \amin$.} This implies that the length of the stream is similar to this semi-major axis with $l(\amin) \approx \amin$. These two estimates of the stream length can be combined to prove the quadratic dependence of the length with distance since $l(\amin)/l(\rt) \approx (\amin/\rt)^2$ using equations \eqref{eq:tidal-radius} and \eqref{eq:semi-major-axis}. The evolution of the stream length can also be expressed in terms of the time $t$ since disruption as $l \propto t^{4/3}$ making use of the dependence $R \propto t^{2/3}$ valid for a near-parabolic orbit. After a time $t \gtrsim \tmin$, the trajectories of most gas elements start to significantly deviate from a parabolic orbit and the above scaling becomes invalid. Once they have escaped the gravity of the black hole, the unbound stream elements move at a constant speed such that their distance from the compact object evolves as $R\propto t$. Because the marginally-bound gas reaches a velocity close to zero, the length of this part of the stream follows the same scaling with $l \propto R \propto t$ \citep{coughlin2016-structure}.

\begin{figure}
\centering
\includegraphics[width=0.9\textwidth]{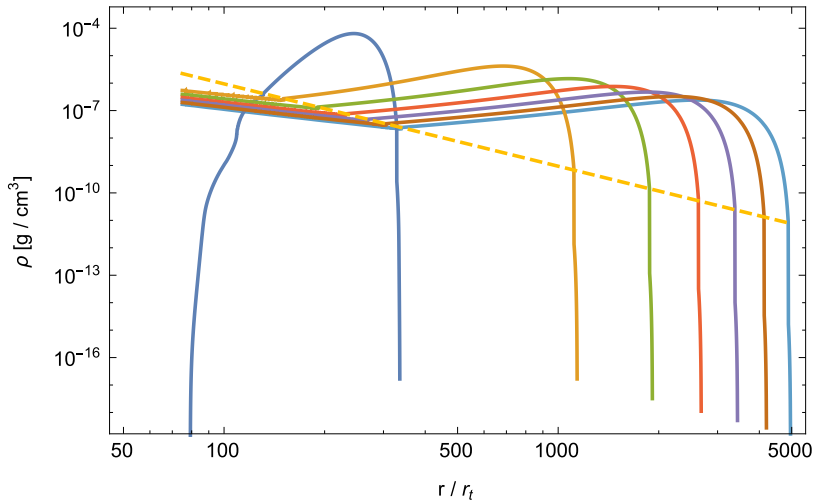}
\caption{Stream density as function of distance from the black hole at different times starting from $t = 32 \days$ (leftmost, dark blue line) and ending at $t = 620 \days$ (rightmost, light blue line) with an interval of $98 \days$ between them \citep{coughlin2016-structure}. The star has a solar mass and radius and is disrupted by a black hole of mass $\mh = 10^6 \msun$ with a penetration factor of $\beta =1$. The gas evolution is assumed to be adiabatic with $\gamma = 5/3$. The yellow dashed line shows the critical density $\rho_{\rm c} = \mh/R^3$ that separates between the self-gravity and shear-dominated regimes.}
\label{fig:stream-density}
\end{figure}

We now focus on the thickness of the gas distribution. Similarly to the above calculation, it is possible to determine the evolution of the width $H$ of a stream element containing a fixed amount of mass as it orbits the black hole. While the densest stream elements have their transverse motion specified by hydrostatic equilibrium between self-gravity and gas pressure, the ones with lower densities have their width set by the shear induced by the tidal force from the black hole. The critical density separating these two regimes can be found by comparing the specific tidal and self-gravity forces acting on the stream that are given by $f_{\rm tid} \approx G \mh H/ R^3$ and $f_{\rm sg} \approx G m /(H l) \approx G \rho H$, respectively. To obtain the latter expression, the Gauss's theorem has been used as well as the relation $m \approx \rho H^2 l$ between the mass $m$ and density $\rho$ of a stream element that exploits its cylindrical geometry. Therefore, an element has its width set by hydrostatic equilibrium with $f_{\rm sg} > f_{\rm tid}$ if $\rho > \rho_{\rm c} \equiv \mh/R^3$ where $\rho_{\rm c}$ denotes the critical density. The next step is to determine the scaling that the stream width obeys in the self-gravity and shear-dominated regimes corresponding to $\rho > \rho_{\rm c}$ and $\rho < \rho_{\rm c}$, respectively. When the tidal force dominates, the width of the stream evolves according to the homologous scaling $H \propto R$. This is because the gas parcels on its cylindrical surface experience a gravitational attraction directed in the radial direction that causes a compression of the stream element as it moves inwards. In the other regime, hydrostatic equilibrium instead imposes that the self-gravity force $f_{\rm sg} \approx G \rho H$ follows the gas pressure force $f_{\rm gas} = \nabla P/\rho \approx P/(\rho H)$, where $P$ denotes the gas pressure that we assume to evolve according to the polytropic relation $P \propto \rho^{\gamma}$. Imposing $f_{\rm gas} \propto f_{\rm sg}$ then leads after some algebra to the relation $H \propto l^{(2-\gamma)/2(\gamma-1)}$. When $t\lesssim \tmin$, the stream length follows equation \eqref{eq:length}, which yields
\begin{equation}
H \propto R^{\frac{2-\gamma}{\gamma-1}}.
\label{eq:width}
\end{equation}
Assuming an adiabatic evolution with $\gamma = 5/3$, the width scales as $H \propto R^{1/2} \propto t^{1/3}$ that corresponds to a density evolution as $\rho \propto (H^2 l)^{-1} \propto R^{-3} \propto t^{-2}$. The unbound elements follow $l \propto R$ at late times $t\gtrsim \tmin$ such that $H \propto R^{(2-\gamma)/2(\gamma-1)} \propto R^{1/4} \propto t^{1/4}$. This latter scaling was found first by \citet{kochanek1994} who assumed the relation $l \propto R$ in the stream. It was later revised to $H \propto R^{1/2}$ at early times by \citet{coughlin2016-structure} that makes use of the correct length evolution. Importantly, these scalings show that self-gravity confines the width of the stream such that it evolves slower with distance from the black hole than for the homologous scaling imposed by the tidal force.

\begin{figure}
\centering
\includegraphics[width=0.9\textwidth]{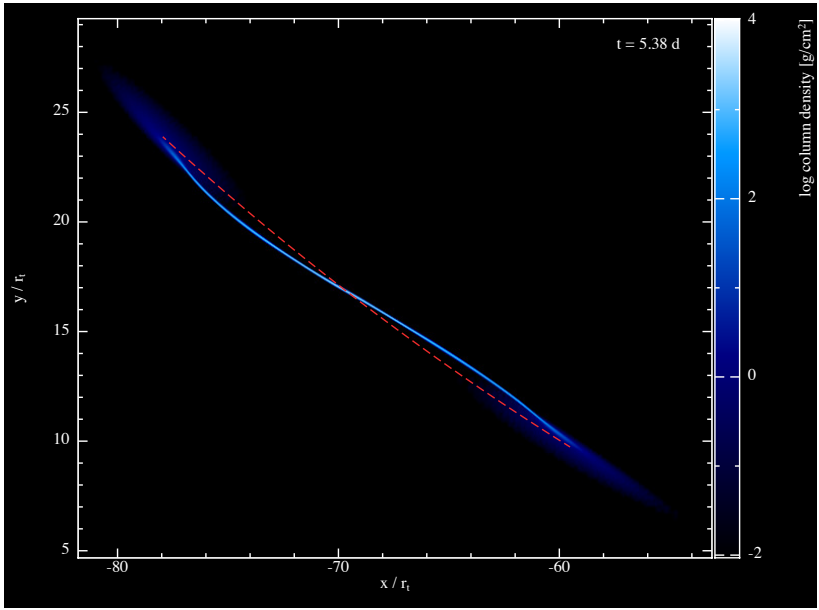}
\caption{Column density of the stream of debris a time of $t = 5.38 \days$ after the disruption of a star of solar mass and radius by a black hole of mass $\mh = 10^6 \msun$ with a penetration factor of $\beta =1$ \citep{coughlin2015-variability}.}
\label{fig:stream-profile}
\end{figure}

For grazing encounters with $\beta \approx 1$, the stellar disruption results in a large fraction of the stream being dominated by self-gravity. This can for example be seen from Fig. \ref{fig:stream-density} that shows stream density profiles at different times with solid lines of different colours according to the semi-analytical model of \cite{coughlin2016-structure}. It considers the disruption of a solar-type star by a black hole of mass $\mh = 10^6 \msun$ with a penetration factor $\beta =1$, additionally assuming an adiabatic gas evolution with $\gamma = 5/3$. Most of the stream elements have a density above the dashed yellow line representing the critical value $\rho_{\rm c} = \mh/R^3$, meaning that they are in hydrostatic equilibrium. This implies that the majority of the gas distribution is able to remain thin thanks to the confinement by self-gravity with a typical width given at $R \approx \amin$ by $H/\rstar \approx (\amin/\rt)^{1/2} \approx (\mh/\mstar)^{1/6} \lesssim 10$ according to the scaling of equation \eqref{eq:width}. Shear-dominated elements are only present at the extremities of the stream due to the lower densities inherited from the original stellar density profile.\footnote{Remarkably, the width evolution becomes homologous with $H\propto R$ near the black hole due to the increased tidal force that corresponds to an up-turn in the density profile seen at $R \lesssim 200 \rt$ in Fig. \ref{fig:stream-density}.} Numerical investigations of the stream evolution by \citet{guillochon2014-10jh} and \citet{coughlin2016-structure} confirm these expectations. It can for example be seen from Fig. \ref{fig:stream-profile} that shows the column density inside the stream obtained from a simulation that adopts the same parameters as the above semi-analytical study. The largest densities are reached in the central parts where the gas distribution remains thin owing to its self-gravity. Instead, the lower densities near the two ends of the stream imply that these regions are shear-dominated and therefore more extended.

Deep disruptions with $\beta \gtrsim 3$ result in heating since the star gets compressed during its disruption, as discussed in the \disrupchap. The ensuing expansion can cause most of the gas to have their densities decreased to $\rho < \rho_{\rm c}$. In this situation, one would expect that the stream is entirely shear-dominated with a width that evolves homologously until late times \citep{kochanek1994,coughlin2016-structure}. This scenario is favoured by the fact that the density evolves at early times as $\rho \propto (H^2 l)^{-1} \propto R^{-4}$ in this regime such that the condition $\rho < \rho_{\rm  c} = \mh/R^3$ remains satisfied if it initially is. However, recent simulations by \citet{Steinberg2019} considering $\beta \geq 5$ find that a significant fraction of the stream is nevertheless able to recollapse under its self-gravity, possibly due to the presence of a weak caustic in the in-plane motion of the debris \citep{coughlin2016-pancakes}. Although it contains more shear-dominated matter than for a grazing encounter, the resulting gas distribution remains therefore qualitatively similar to that described above for $\beta \approx 1$.

\subsection{Additional physics}
\label{sec:additional}

\subsubsection{Gravitational fragmentation}

As the stream expands, adiabatic cooling can trigger gravitational fragmentation with individual clumps forming within it. This phenomenon was first found in simulations by \citet{coughlin2015-variability} that consider a tidal disruption with $\beta = 1$ assuming an adiabatic evolution with $\gamma = 5/3$. The perturbation triggering fragmentation in this work likely has a numerical origin and the exact outcome depends the resolution used. \citet{coughlin2016-pancakes} identified a possible physical cause for fragmentation, however, which is produced by the in-plane compression of the star as it is being disrupted by the black hole. This compression is due to the fact that, when the star is at the pericenter of its trajectory, the gas that has already passed this location decelerates while that lagging behind keeps accelerating. This results in a density increase happening a few hours after the disruption followed by oscillations as the stream settles back to hydrostatic equilibrium. The associated perturbation was proposed to destabilize the stream and trigger gravitational fragmentation but additional mechanisms likely exist such as the interaction with the surrounding medium discussed in Section \ref{sec:interaction}.

The conditions for gravitational fragmentation have been examined more precisely by \citet{coughlin2016-structure}. It requires that the free-fall time $t_{\rm ff} \approx (\rho G)^{-1/2}$ is locally shorter than the dynamical time $t_{\rm dyn} \approx (G \mh/R^3)^{-1/2}$ necessary for the stream to get stretched in the radial direction. The condition $t_{\rm ff} < t_{\rm dyn}$ then translates to $\rho \gtrsim \mh/R^3 = \rho_{\rm c}$, which is identical to that derived in Section \ref{sec:basic} for a stream element to be in hydrostatic equilibrium. In this regime, the density scales as $\rho \propto R^{-2/(\gamma -1)}$, implying that the condition for fragmentation remains valid at later times as long as $\gamma \geq 5/3$. This analysis suggests that clump formation is only possible for adiabatic exponents larger than this critical value and occurs faster when $\gamma$ is increased. Numerical simulations by \citet{coughlin2016-pancakes} confirm this trend by studying the stream evolution for different values of the adiabatic exponent. In reality, $\gamma > 5/3$ requires that the stream cools non-adiabatically, which can for instance be achieved through recombination when the gas becomes optically thin (see Section \ref{sec:recombination}).

The properties of these clumps such as their preferred location along the stream are determined by the gravitational instability producing them, whose development has been studied in details by \citet{coughlin2020}. Due to the local density enhancements they produce, the fragments that belong to the bound part of the debris introduce variability in the fallback rate \citep{coughlin2015-variability}. The resulting emission could also be affected by the presence of these clumps, although including the impact of pressure delays their formation such that they may fall back once the emission has already faded below detection limits \citep{sacchi2020}. The fragments forming in the unbound part of the stream instead escape the black hole, which constitutes a new kind of high-velocity objects \citep{coughlin2016-pancakes}. Their spatial distribution has been studied more in detail by \cite{girma2018} who find that $\sim 10^7$ of them have been launched from Sgr A$^{\star}$ within the life of our galaxy.

\subsubsection{Recombination}
\label{sec:recombination}

Immediately after the disruption, the gas is fully ionized with a temperature of $T_{\star} \approx G \mstar m_{\rm p}/ (k_{\rm B} \rstar) = 2 \times 10^7 \K$ similar to that necessary to ensure hydrostatic equilibrium inside the original star, as shown in the \disrupchap. Here, $k_{\rm B}$ denotes the Boltzmann constant while $m_{\rm p}$ is the proton mass. As it subsequently evolves around the black hole, this debris cools
adiabatically, as its large optical depths prevent the heat from being promptly radiated. The temperature then evolves as $T \propto \rho^{2/3} \propto t^{-4/3}$ using the scaling $\rho \propto t^{-2}$ valid at early times assuming hydrostatic equilibrium (Section \ref{sec:basic}). This leads to $T \approx T_{\star} (t/t_{\rm str})^{-4/3}$, where $t_{\rm str} \approx (\rt^3/G \mh)^{1/2} \approx 1 \hours$ is the duration required for the stream to stretch by a significant amount following the disruption that is similar to the dynamical timescale near the tidal radius or, equivalently, the stellar dynamical timescale according to equation \eqref{eq:tidal-radius}. Cooling continues until the temperature drops to a value of $T_{\rm rec} \approx 10^4 \K$, at which hydrogen recombines. Setting $T = T_{\rm rec}$, this occurs after a time $t_{\rm rec } \approx t_{\rm str} (T_{\star}/T_{\rm rec})^{3/4} \approx 10 \days$ \citep{coughlin2016-structure}.

The effect of recombination on the stream depends on its optical depth \citep{kochanek1994,coughlin2016-structure,kasen2010}. By the time it occurs, the majority of the gas is still very optically thick, implying that this debris gets heated by the additional energy injection. During this process, the stream is imposed to evolve at a roughly constant temperature and becomes overpressured, causing an increase of its thickness and the end of its confinement by self-gravity. Since the stream outer layers are less dense, they are able to radiatively cool and become neutral that causes a decrease of the opacity since fewer free electrons are present. This process drives a transparency wave that propagates from the outer regions through the entire stream in a few days while producing an optical transient emission with a luminosity of $L_{\rm rec} \lesssim 10^{41} \ergpers$, as estimated by \citet{kasen2010}. However, these values may be overestimated because this work assumes a free expansion of the stream after disruption that increases the emitting area compared to the slower thickening found in more recent investigations due to the confining effect of self-gravity.

\subsubsection{Magnetic fields}
\label{magnetic}

The magnetic field of the star gets transferred to the debris upon disruption and is subsequently carried by the stream in its evolution around the black hole. Its magnetic strength is then entirely specified by the stream geometry derived in Section \ref{sec:basic} according to magnetic flux conservation. This condition imposes that the magnetic field parallel to the direction of stream elongation evolves as $B_{\parallel} \propto S^{-1}_{\parallel}$ where $S_{\parallel} \approx H^2$ denotes the surface the corresponding field lines go through. This results in $B_{\parallel} \propto R^{-1}$ according to equation \eqref{eq:width} and assuming $\gamma = 5/3$. Similarly, the magnetic field orthogonal to the direction of stream stretching evolves as $B_{\perp} \propto S^{-1}_{\perp}$ with $S_{\perp} \propto H l$ such that $B_{\perp} \propto R^{-5/2}$ additionally using equation \eqref{eq:length}. The fact that the parallel component decreases much slower than the perpendicular one implies that the field lines tend to align along the stream longitudinal direction.

Given this evolution, it is possible to evaluate the relative importance of magnetic and gas pressure inside the stream. Assuming that the field lines are all parallel to the direction of stream elongation, magnetic pressure follows $P_{\rm mag} \approx B^2_{\parallel} \propto R^{-2}$ while the gas pressure scales as $P_{\rm gas} \propto \rho^{5/3}  \propto R^{-5}$ for an adiabatic evolution. This shows that the ratio of magnetic to gas pressure increases as $P_{\rm mag}/P_{\rm gas} \propto R^3 \propto t^2$. Magnetic pressure takes over gas pressure with $P_{\rm mag}/P_{\rm gas} = 1$ after a time given by $t_{\rm mag} \approx t_{\rm str} \beta^{1/2}_{\star}$ where $t_{\rm str} \approx 1 \hours$ has been defined in Section \ref{sec:recombination} while $\beta_{\star}$ denotes the initial plasma beta, defined as the ratio of gas to magnetic pressure inside the original star. This parameter can be evaluated as $\beta_{\star} \approx 10^{16} (B_{\star}/1 G)^{-2}$ denoting by $B_{\star}$ the stellar magnetic field. Decreasing $t_{\rm mag}$ to about a month such that it becomes of order the orbital period $\tmin$ of the most bound debris defined in equation \eqref{eq:fallback-time} therefore requires $B_{\star} \gtrsim 10^5 G$ that corresponds to a highly-magnetized star. In this situation, the stream gets dominated by magnetic pressure that can cause a fast increase of its width and the end of confinement by self-gravity \citep{guillochon2017-magnetic,bonnerot2017-magnetic}.

\subsubsection{Stream-disc interaction}
\label{sec:interaction}

So far, we have assumed that the stream evolves in isolation but the debris may in fact be affected by the ambient gaseous medium present around the black hole. A possible interaction relates to the Kelvin-Helmholtz instability that can develop at the interface between the debris and the surrounding matter, resulting in an efficient mixing of these two fluids. A given stream element is affected by this process if the instability has fully developed by the time the corresponding debris comes back to pericenter. Using this criterion, \cite{bonnerot2016-kh} finds that the entire stream can be impacted for the disruptions of red giants by black holes of mass $\mh \gtrsim 10^9 \msun$. In this situation, most of the debris likely gets mixed with the ambient medium that can result in a sub-luminous event. Although this type of disruption is promising to probe the high-end of the black hole mass function \citep{macleod2014}, the above process may significantly complicate their detection. If the galaxy contains a denser disc, tidal streams produced by main sequence stars can also be affected by the instability. This possibility has been put forward to explain the sudden drop in luminosity detected from some events \citep{kathirgamaraju2017} and could be involved in the capture of the disc magnetic field by the debris stream \citep{kelley2014}.

Additional effects can take place if a TDE happens in an active galactic nucleus that contains a dense accretion flow.\footnote{Note that the rate of tidal disruptions in such systems may be increased due to the modification of the gravitational potential by the mass of the disc \citep{karas2007}, as discussed in more details in the \ratechap{}.} Using hydrodynamical simulations, \citet{chan2019} studied the interaction of a debris stream with such a disc and find two different outcomes depending on the density of the incoming stellar matter. For low densities, the stream deposits most of its kinetic energy in the disc to get promptly absorbed by it. A stream of higher density can instead pierce through the disc while significantly perturbing the inflowing matter that can result in a fraction of the stream getting unbound and a depletion of the inner part of the accretion flow.

\section{Disc formation} 

\label{sec:disc-formation}

After a revolution around the black hole, the bound part of the stream comes back to the stellar pericenter. We are interested here in the fate of this gas and how it evolves to form an accretion flow. A simple although unrealistic scenario consists in the immediate formation of a circular disc from the infalling debris near the circularization radius $R_{\rm circ} \approx 2 \rp$ that corresponds to the lowest energy state for the gas if each fluid element conserves its specific angular momentum.\footnote{If all the debris have the same specific angular momentum as that of the star on its original parabolic orbit, $l_{\star} = \sqrt{2 G \mh \rp}$, the lowest energy state is a circular orbit at a distance $R_{\rm circ} = l^2_{\star}/G \mh = 2 \rp$ from the black hole. Although the most bound part of the stream is on an elliptical orbit, this estimate remains valid since its typical eccentricity is $\emin = 0.98 \approx 1$ as shown in equation \eqref{eq:eccentricity}.} In practice, this full circularization of the trajectories requires that a specific energy
\begin{equation}
\Delta \varepsilon_{\rm circ} \approx  \frac{G \mh}{4 \rp},
\label{eq:energy-circular}
\end{equation}
is entirely lost from the system.\footnote{Although the most bound part of the stream comes back to pericenter with an already negative energy given by equation \eqref{eq:energy-spread}, it is much larger than that required to reach complete circularization, since $\Delta \varepsilon/\Delta \varepsilon_{\rm circ} \approx \rp/\amin \ll 1$, and therefore $\Delta \varepsilon$ is irrelevant in the computation of the energy loss necessary to reach this configuration.} This picture was first described in early analytical works by \citet{rees1988} and \citet{ulmer1999}, which pointed out that the timescale of disc formation is uncertain but may be of order the fallback time of equation \eqref{eq:fallback-time}. They also predicted that the disc would be located near the circularization radius $R_{\rm circ}$ and have a thick vertical profile due to the impact of radiation pressure on the gas. Since these pioneering works, multiple investigations have been made that aim at improving this original picture of disc formation. An exhaustive list of these works is included in Table \ref{tab:simulations} along with the choice of parameters made and the method used. These recent studies showed that the early interactions experienced by the returning stream typically dissipate an energy much lower than $\Delta \varepsilon_{\rm circ}$, implying that disc formation could require more than a fallback time to complete. The resulting gas distribution is also found to typically reach distances much larger than the circularization radius due to angular momentum redistribution between interacting fluid elements.\footnote{There is also observational evidence supporting the notion that disc formation occurs differently from what early works assumed. For example, a compact accretion disc cannot reproduced the high level of optical emission detected from many events \citep{lodato2010,miller2015-winds}. Additionally, most TDEs have an integrated energy lower than $\mstar \Delta \varepsilon_{\rm circ} \approx 10^{52} \erg $, by at least an order of magnitude. This ``inverse energy crisis'' is discussed in Section \ref{sec:ballistic} along with possible solutions.} Additionally, we emphasize that the gas evolution is expected to be qualitatively different depending on the region of parameter space considered. Although we refer to the gas structure formed from the debris around the black hole as a  ``disc'', it can significantly differ from the simplest disc models \citep[e.g.][]{shakura1973} used in the literature.\footnote{As will be described more precisely in Section \ref{sec:hydrodynamics}, it is for example possible that this disc remains globally eccentric, or inclined with respect to the black hole spin.} Despite recent progress in the understanding of disc formation, it is important to highlight that no clear consensus has so far been reached concerning the hydrodynamics of this process, largely due to numerical limitations that have so far prevented fully realistic global simulations of disc formation. In the following, we attempt to provide a description of the current status of theoretical knowledge, remaining as agnostic as possible while also highlighting (i) the main sources of uncertainties and (ii) the areas where secure conclusions appear possible.

In Section \ref{sec:initial}, we describe the dissipation mechanisms taking place shortly after the return of the stream near the black hole. The numerical challenges inherent to a global simulation of disc formation are discussed in Section \ref{sec:numerical} along with the compromises made so far to alleviate them. Section \ref{sec:hydrodynamics} presents the results of these numerical works that capture the evolution of the gas as the accretion flow assembles. The radiative efficiency of the different shocks involved in this process and the resulting emission are evaluated in Section \ref{sec:shock-driven}. Finally, Section \ref{sec:nascent} describes the consequences of disc formation on the subsequent phase of accretion onto the black hole.

\subsection{Early sources of dissipation}
\label{sec:initial}

We start by discussing dissipation mechanisms taking place shortly after the debris comes back near the black hole, including the nozzle shock and the self-crossing shock. The nozzle shock is due to a strong compression of the stream during pericenter passage that results in an expansion of the gas as it then moves outward. Later on, an intersection of the stream with itself causes a self-crossing shock that can induce a significant modification of the gas trajectories.

\subsubsection{Nozzle shock}
\label{sec:nozzle}

An early source of dissipation occurs during the passage of the stream at pericenter. In this region, the gas trajectories are initially specified by the gravity of the black hole. A consequence is that the fluid elements furthest from the original stellar orbital plane move on inclined orbits that intersect near pericenter, resulting in a compression of the stream in the vertical direction. This effect is analogous to the compression experienced by a star if it is disrupted on a plunging trajectory with $\beta >1$. Here, the stream is always inside its own tidal radius, with an effective penetration factor $\beta_{\rm s} \gg \beta$ since its density has decreased compared to the stellar value by several orders of magnitude (see Section \ref{sec:stream-evolution}). The compression results in a nozzle shock, during which the kinetic energy associated with vertical gas motion near pericenter gets dissipated. Note that compression also occurs within the orbital plane due to a convergence of the trajectories along this direction. However, this effect is not guaranteed to result in an intersection of the fluid elements, which likely reduces its impact on the gas evolution compared to that of vertical compression.

As discussed in the \disrupchap, the vertical velocity associated to the stellar compression is given approximately by $v_{\rm z} \approx \alpha \, v_{\rm p} \approx \beta c_{\rm s, \star}$ where $c_{\rm s, \star} =\sqrt{G\mstar/\rstar}$ denotes the stellar sound speed. Here, $v_{\rm p} = (G \mh/\rp)^{1/2}$ is a characteristic velocity near pericenter while $\alpha = \rstar / \sqrt{\rt \rp}$ denotes the inclination angle between the orbital plane of the compressed gas and that of the stellar center of mass \citep{carter1982}. As pointed out by \citet{guillochon2014-10jh}, this expression can be extrapolated to study the compression of the returning debris, for which the vertical velocity becomes $v_{\rm z,s} \approx \beta_{\rm s} c_{\rm s,s}$ where $c_{\rm s,s}$ represents the sound speed inside the stream. Denoting by $\rho_{\star}$ and $\rho_{\rm s}$ the stellar and stream densities, the ratio of penetration factors is given by $\beta_{\rm s}/\beta \approx (\rho_{\star}/\rho_{\rm s})^{1/3}$ since the tidal radius scales as $\rt \propto \rho^{-1/3}$ while the pericenter distance is unchanged. Similarly, the sound speed ratio can be evaluated as $c_{\rm s,s}/ c_{\rm s,\star} \approx (\rho_{\star}/\rho_{\rm s})^{-1/3}$ since $c_{\rm s} \propto (P/\rho)^{1/2} \propto \rho^{1/3}$ under the legitimate assumption of an adiabatic evolution for the debris with an adiabatic exponent $\gamma = 5/3$. The density ratio therefore cancels out such that $v_{\rm z,s} \approx v_{\rm z}$.\footnote{This relation can also be understood from the fact that the stream has a width similar to the original stellar radius when it comes back near pericenter. Because the compression takes place near the tidal radius, it  occurs on the dynamical timescale of the star, which leads to $v_{\rm z,s} \approx \sqrt{G\mstar/\rstar}$.}

The nozzle shock dissipates a large fraction of the specific kinetic energy associated with the vertical motion of the stream,
\begin{equation}
\Delta \varepsilon_{\rm no} = v^2_{\rm z,s} \approx \beta^2 \frac{G \mstar}{\rstar},
\label{eq:nozzle}
\end{equation}
where we have used the above estimates for the vertical speed of $v_{\rm z,s} = \beta  \sqrt{G\mstar/\rstar}$. It follows that $\Delta \varepsilon_{\rm no}/\Delta \varepsilon_{\rm circ} \approx \alpha^2 \approx 10^{-4} \beta (\mh/10^6 \mstar)^{-2/3}$, making use of equation \eqref{eq:energy-circular}. This calculation suggests that the dissipation provided by the nozzle shock is negligible compared to that necessary to completely circularize the orbits except for very low black hole masses $\mh \lesssim 10^4 \msun$, where $\Delta \varepsilon_{\rm no} \gtrsim 10^{-1} \Delta \varepsilon_{\rm circ}$ \citep{guillochon2014-10jh}. The internal energy injected in the shocked gas can nevertheless cause the stream to expand significantly after leaving pericenter, an effect captured by several simulations described in Section \ref{sec:hydrodynamics}, and especially strong for those considering an intermediate-mass black hole \citep{ramirez-ruiz2009,shiokawa2015,guillochon2014-10jh} or a large penetration factor \citep{sadowski2016}.

So far, the nozzle shock has not been studied in a systematic fashion for typical parameters of the problem and our understanding of it is therefore limited. In particular, it is unclear whether the resulting gas expansion significantly modifies the stream properties compared to that before pericenter passage. As discussed more in details in Section \ref{sec:self-crossing}, this effect can have consequences on the later evolution of the gas. Another important uncertainty concerns the exact amount of energy dissipated by this interaction. One possibility is that a fraction of the kinetic energy associated with the orbital motion is dissipated during this process, in addition to the much lower reservoir of vertical kinetic energy we have discussed so far. Due to the strong shearing expected at pericenter between neighbouring fluid elements, additional dissipation may be caused by effective viscosity, potentially provided by magnetic fields. Alternatively, additional shocks can take place if more intersections of the trajectories occur, which could result from general relativistic effects such as apsidal and nodal precessions, the latter being due to the black hole spin. As a result, the level of heating would be higher than that predicted from equation \eqref{eq:nozzle}, causing the gas to expand faster from pericenter.

\subsubsection{Self-crossing shock }
\label{sec:self-crossing}

Another early dissipation mechanism consists of a self-crossing shock caused by the intersection between the part of the stream that has passed pericenter and the debris still approaching the black hole. The way this collision takes place depends on the following physical effects: (i) relativistic apsidal precession, which modifies the stream trajectory near pericenter to put it on a collision course with the infalling gas, (ii) the expansion resulting from the nozzle shock described in Section \ref{sec:nozzle}, which leads to an increase of the stream width after pericenter passage and (iii) nodal precession induced by the black hole spin, which changes the orbital plane of the gas during pericenter passage. The works that specifically studied this process are included in Table \ref{tab:simulations} for entries whose first letter in the the last column indicating the method used is either `L' or `B'. We start by evaluating the influence of these three mechanisms on the self-crossing shock and then describe the gas evolution during the collision for different regimes specified by the relative importance of these effects.

\begin{figure}
\centering
\includegraphics[width=\textwidth]{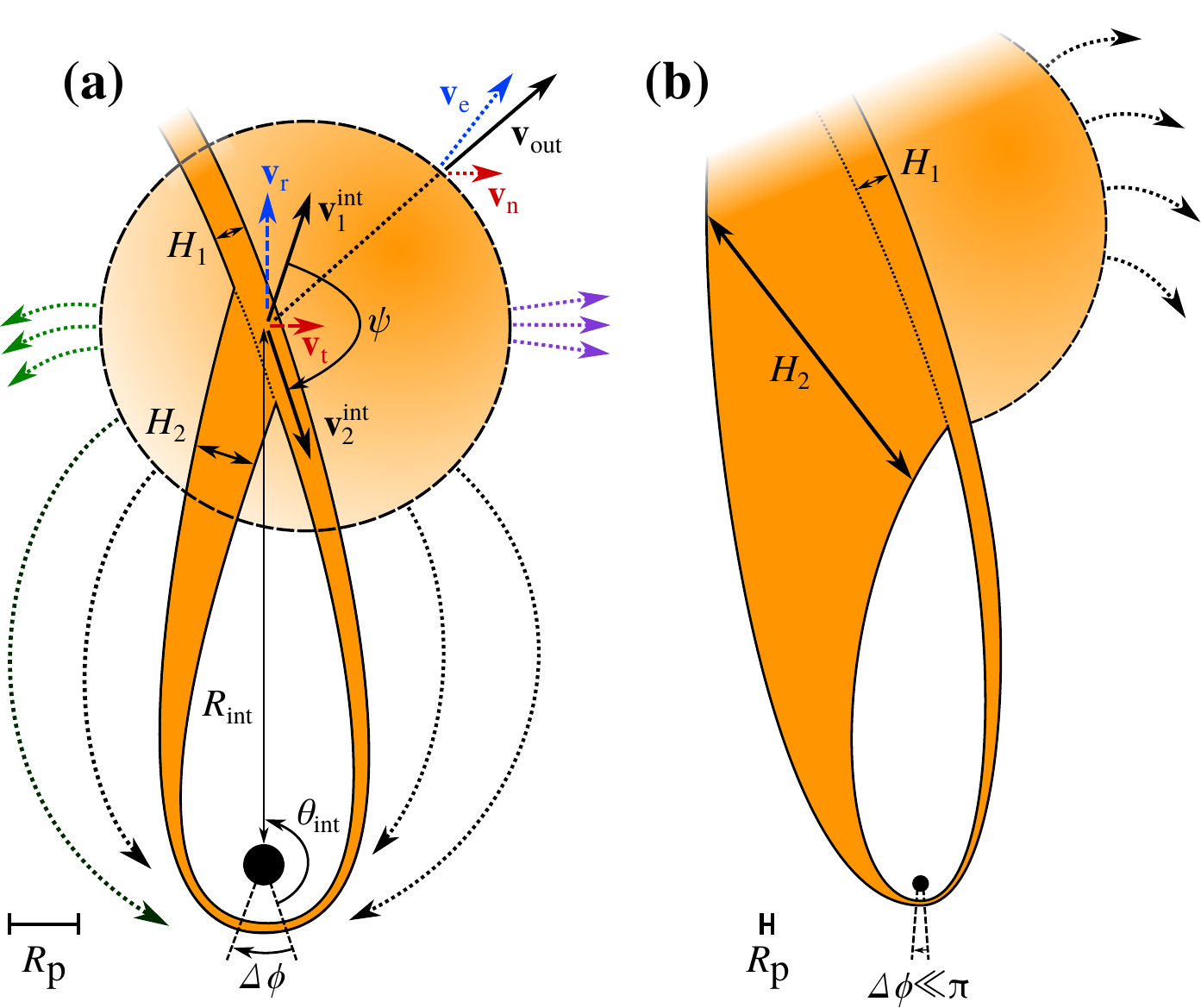}
\caption{Sketches representing the self-crossing shock caused by an intersection of the stream with itself, and the subsequent gas evolution in two different regimes. In the left panel (a), the collision is dominated by relativistic apsidal precession, which makes the tip of the stream precess by an angle $\Delta \phi$ during pericenter passage. The resulting intersection takes place at a distance $\rint$ from the black hole that corresponds to a true anomaly $\theta_{\rm int} = \pi - \Delta \phi/2$ for the trajectories of the stream components that are assumed to have retained similar widths $H_1 \approx H_2 \ll \rint$. Their velocities $\mathbf{v}^{\rm int}_1$ and $\mathbf{v}^{\rm int}_2$ are of the same magnitude but are misaligned by the collision angle $\psi$, such that they have identical tangential components $\mathbf{v}_{\rm t}$ but opposite radial ones $\pm \mathbf{v}_{\rm r}$. If the widths of the colliding gas are identical with $H_1 = H_2$ with no vertical offset $\Delta z =0$ between them, all their mass passes through the self-crossing shock, resulting in the dissipation of the kinetic energy associated with the gas radial motion. This causes the formation of an outflow (orange sphere delimited by black dashes) driven by radiation pressure, whose velocity can be decomposed into a net and expansion component as $\mathbf{v}_{\rm out} = \mathbf{v}_{\rm e} + \mathbf{v}_{\rm n}$ where $|\mathbf{v}_{\rm e}| \approx |\mathbf{v}_{\rm r}|$ and $\mathbf{v}_{\rm n} \approx \mathbf{v}_{\rm t}$. Gas expanding in the direction of its net motion (dotted purple arrows) can get unbound from the black hole while that moving in the opposite way (green dotted arrows) can acquire a direction of motion opposite to that of the original star. If the stream components have a slight width difference with $H_1 \lesssim H_2$ or are weakly misaligned with $\Delta z \ll H_1 + H_2$, a small fraction of the gas can avoid the intersection while the rest experiences a collision qualitatively similar to that described above. In the right panel (b), we consider a different limit, wherein the expansion following the nozzle shock produces very different widths for the two stream components, namely $H_1 \ll H_2$. As a result, intersections occur across a broad range of radii even if relativistic precession is negligible ($\Delta \phi \ll \pi$). Due to the large density difference between the two colliding components, the infalling gas can pass through the outgoing debris with a largely unaffected trajectory. It is possible that a bow shock forms that heats the outward-moving gas causing its mild expansion (orange half-sphere delimited by black dashes) following the intersection. In both regimes, the gas that remains bound after the self-crossing shock eventually returns near the black hole (black dotted arrows) where it can experience additional interactions.}
\label{fig:sketch}
\end{figure}

\paragraph{(i) Relativistic apsidal precession:} When the stream passes at pericenter, its center of mass precesses by an angle that can be approximated for a non-rotating black hole as \citetext{see equation (10.2) of \citealt{hobson2006}}
\begin{equation}
\Delta \phi \approx 3 \pi \rg/\rp \approx 11.5^{\circ} \beta \left( \frac{\mh}{10^6 \msun} \right)^{2/3},
\label{eq:precession_angle}
\end{equation}
where $\rg = G \mh/c^2$ is the gravitational radius and the approximation $\amin (1-\emin^2) \approx 2 \rp$ has been used. The precession angle decreases or increases compared to equation \eqref{eq:precession_angle} if the black hole has a non-zero spin with angular momentum vector in the same or opposite direction to that of the star's orbit, respectively. Apsidal precession by itself can result in a collision between the part of the stream moving outward and that still approaching the black hole.\footnote{Another source of interactions for the stream involves the sequential tidal disruption of the two components of a binary star by a black hole \citep{bonnerot2019-precursor}, during which the two debris streams can collide with each other due to the difference in their trajectory induced by the previous binary separation.} The self-crossing shock taking place when this mechanism dominates is illustrated in the left panel (a) of Fig. \ref{fig:sketch}. Assuming that the two colliding components remain thin with widths $H_1 \approx H_2 \ll \rint$, the collision can be considered to happen at a single location. Its characteristics can then be derived analytically by considering two Keplerian ellipses of eccentricity $e_{\rm min}$ and pericenter $\rp$ rotated according to the above precession angle, as done by several authors \citep{dai2015,jiang2016,bonnerot2017-stream,lu2020}.\footnote{Note that this treatment is approximate due to the difference in orbital energy between the two gas components involved in the collision. In reality, the part of the stream moving away from the black hole is more bound than the approaching one with a lower apocenter distance.} The resulting collision occurs at a true anomaly $\theta_{\rm int} = \pi - \Delta \phi/2$ that corresponds to an intersection radius of
\begin{equation}
R_{\rm int} = \frac{1}{\beta} \frac{ \rt (1+e_{\rm min})}{1-e_{\rm min} \cos (\Delta \phi/2)},
\label{eq:intersection_radius}
\end{equation}
which decreases monotonically from the apocenter of the most bound debris to its pericenter as the relativistic apsidal angle $\Delta \phi$ increases. The two stream components collide with velocity vectors $\mathbf{v}^{\rm int}_1$ and $\mathbf{v}^{\rm int}_2$ of the same magnitude $v_{\rm int} \equiv |\mathbf{v}^{\rm int}_1| = |\mathbf{v}^{\rm int}_2|$ but inclined by the collision angle $\psi$ given by
\begin{equation}
\cos \psi = \frac{1-2 e_{\rm min} \cos (\Delta \phi/2) + e^2_{\rm min} \cos \Delta \phi}{1-2 e_{\rm min} \cos (\Delta \phi/2) + e^2_{\rm min} },
\label{eq:collision_angle}
\end{equation}
which can be obtained from $\cos \psi = \mathbf{v}^{\rm int}_1 \cdot \mathbf{v}^{\rm int}_2/v^2_{\rm int}$ and expressing the two velocities as a function of $\rint$. This angle is small only when apsidal precession is very weak, corresponding to low-mass black holes, i.e. $\mh \lesssim 10^5 \msun$ according to equation \eqref{eq:precession_angle}. Otherwise, the velocity vectors are significantly misaligned, with directions close to being completely opposite with $\psi \approx \pi$. In the typical situation where $\mh \approx 10^6 \msun$ and $\beta \approx 1$, apsidal precession is relatively weak such that the collision happens near apocenter with $\rint \approx 2\, \amin$ according to equation \eqref{eq:intersection_radius}. For the two stream components to collide instead near pericenter with $\rint \approx \rp$, apsidal precession must be stronger, requiring that the black hole mass increases to $\mh \gtrsim 10^7 \msun$ if $\beta \approx 1$ or the penetration factor to $\beta \gtrsim 5$ if $\mh \approx 10^6 \msun$.

\begin{figure}
\centering
\includegraphics[width=0.29\textwidth]{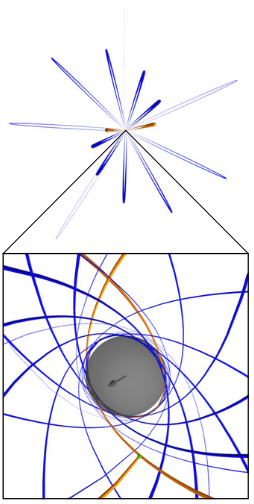}
\hspace{0.2cm}
\includegraphics[width=0.67\textwidth]{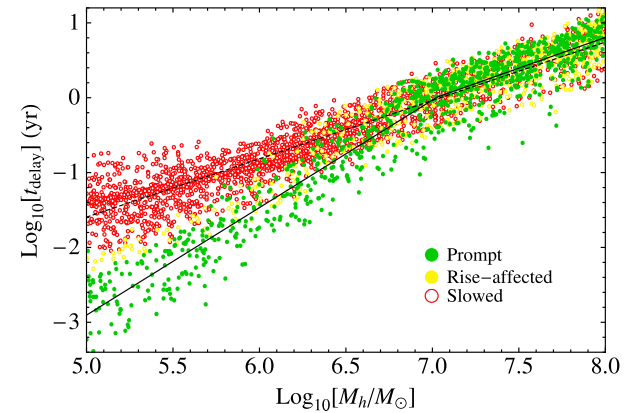}
\caption{Left panel: Trajectory of the stream of debris around a spinning black hole (grey sphere) obtained from the semi-analytical model of \cite{guillochon2015}. Because of Lense-Thirring precession, the stream experiences several windings (blue curves) before eventually colliding, as shown with the orange curves that intersect at the green point. Right panel: Time $t_{\rm delay}$ elapsed between the passage of the star at pericenter and the intersection of the debris stream as a function of black hole mass. Each point corresponds to a different set of parameters assumed for the above process with the colours related to how fast the resulting disc viscously accretes with respect to the time of peak fallback rate.}
\label{fig:guillochon}
\end{figure}

\paragraph{(ii) Expansion from the nozzle shock:} As described in Section \ref{sec:nozzle}, dissipation takes place during the strong compression of the stream at pericenter, leading to gas expansion as it then moves away from the black hole. As a result, this part of the stream can be thicker than that still moving inwards, so that the nozzle shock induces an asymmetry between these two components. Due to our limited understanding of the nozzle shock, its impact on the stream width past pericenter has so far not been precisely quantified. As we explained, the amount of dissipation during this interaction is uncertain, being potentially much larger than current analytical estimates. Additionally, the direction along which the gas preferentially bounces following the compression is largely unknown. It is therefore unclear whether the shocked part of the stream mainly expands within the orbital plane or perpendicular to it. This effect is particularly important since it determines the geometry of the outgoing stream component that may be involved in the self-crossing shock. To take this ambiguity into account, we allow this part of the stream to become thicker than the infalling matter, with $H_2\geq H_1$ as a result of the nozzle shock.  The width\footnote{Although for simplicity this description involves a single width $H_2$ for the outgoing stream component, it is not guaranteed that the gas distribution remains cylindrical following the expansion.} $H_2$ can in principle take a wide range of possible values, from $H_2 = H_1$ up to a size comparable to the intersection radius, i.e. $H_2 \approx \rint$. The left panel (a) of Fig. \ref{fig:sketch} illustrates the regime where the self-crossing shock is still mainly determined by relativistic apsidal precession, but involves stream components of slightly different widths with $H_2 \gtrsim H_1$. In the right panel (b) of Fig. \ref{fig:sketch}, relativistic precession is negligible and the streams have very different widths, with $H_1 \ll H_2 \approx \rint = 2\, \amin$. The large thickness of the outgoing gas implies that the intersection takes place at a wide range of distances from the black hole. These interactions are indirectly caused by the nozzle shock that deflects a fraction of the gas in the direction of the infalling debris, even in the absence of relativistic precession. A self-crossing shock induced by this effective precession may represent an important dissipation mechanism when relativistic precession is very weak. This scenario seems particularly applicable to black holes with masses $\mh \lesssim 10^5 \mh$, for which the dynamical impact of the nozzle shock is additionally enhanced, as predicted by the estimates of Section \ref{sec:nozzle}. A few simulations considering intermediate-mass black holes \citep[][]{ramirez-ruiz2009,guillochon2014-10jh,shiokawa2015} have captured this effect but its role in the disc formation process has not been further investigated in a systematic way.

\paragraph{(iii) Spin-induced nodal precession:} If the black hole has a non-zero spin, the stream experiences nodal precession (dominated by the Lense-Thirring effect) near pericenter that modifies its orbital plane. As first pointed out by \citet{cannizzo1990}, this effect produces a vertical offset $\Delta z$ between the infalling and outgoing debris at the intersection radius where they would otherwise collide. If $\Delta z < H_1 + H_2$, the self-crossing shock still occurs, but the two stream components are misaligned. The interaction is entirely avoided if $\Delta z > H_1 + H_2$: in this case, the streams are unable to impact one another, and keep following their trajectories unimpeded. \citet{jiang2016} carried out the following analytical estimate to determine whether this initial self-crossing shock is avoided. Nodal precession causes the gas angular momentum vector to precess around the black hole spin direction by an angle $\Delta \Omega \approx 4 \pi \, 2^{-3/2} a  \sin i  \, (\rg/\rp)^{3/2} \approx 1^{\circ} a \sin i\, \beta^{3/2} (\mh/10^6 \msun)$ \citep{merritt2013}, where $a$ denotes the dimensionless spin parameter and $i$ represents the angle between the black hole spin and the stellar angular momentum that is generally non-zero.\footnote{As explained in the \ratechap{}, victim stars sourced by two-body relaxation approach the black hole with a quasi-isotropic distribution of orientations, although alternative scenarios \citep{Wernke2019} are possible. As a result, the stellar angular momentum is in general not aligned with the black hole spin: $i\neq 0$.} The resulting shift in orbital plane causes the centers of mass of the two gas components to be separated by a distance of $\Delta z \approx \rint \Delta \Omega$ at the intersection radius given by equation \eqref{eq:intersection_radius}. Additionally, the sum of the stream widths at this position is approximated by a single value $H = H_1 + H_2 \approx \rstar \rint/\rp$ that expands homologously (see Section \ref{sec:stream-evolution}) from pericenter where it is assumed to be of a stellar radius. The ratio of vertical offset to stream width is then given by
\begin{equation}
\frac{\Delta z}{H} \approx a \sin i \, \beta^{1/2} \left( \frac{\mh}{10^6 \msun} \right)^{4/3}.
\label{eq:offset}
\end{equation}
A prompt collision between misaligned stream components can take place if $\Delta z < H$, which is satisfied for low-mass and slowly-spinning black holes with $\mh \lesssim 10^6 \msun$ and $a \lesssim 1$ if $\beta \approx 1$ and $i \approx 90^\circ$ according to this estimate. The intersection is instead completely avoided with $\Delta z > H$ if the black hole has a larger mass $\mh \gtrsim 10^7 \msun$ and is maximally-spinning with $a \approx 1$, keeping the other parameters fixed. If this early self-crossing shock does not occur, the stream needs to complete additional revolutions around the black hole before eventually colliding with itself. \cite{dai2013} studied this effect first by integrating the Kerr geodesics followed by the stream. However, the impact of nodal precession was likely overestimated by assuming that the gas is still confined by self-gravity, which is unlikely to be valid near the black hole. A more extensive investigation was done by \citet{guillochon2015} using a more complex, semi-analytical geodesic model that assumes that the stream width evolves homologously ($H\propto R$), and also includes the additional expansion induced by the nozzle shock at each pericenter passage. In the regime where prompt self-intersection fails, streams orbiting the black hole will therefore become progressively thicker until two different windings eventually intersect with $\Delta z < H_1 + H_2$, leading to a self-crossing shock. An example of stream evolution obtained from this work is shown in the left panel of Fig. \ref{fig:guillochon}, while the right panel displays the time $t_{\rm delay}$ between the stellar disruption and the first successful collision as a function of black hole mass. This time delay is typically several orbital periods and increases with black hole mass to reach of order a year for $\mh = 10^7 \msun$. A delayed intersection is also more likely to happen between two successive windings, according to this study. In this case, the trajectories of the colliding debris may be close to that of the initial self-crossing shock shown in the left panel (a) of Fig. \ref{fig:sketch}. However, the stream components may have a residual misalignment and widths larger than that of the initially returning debris due to the cumulative effect of successive nozzle shocks. Additionally, the cold streams present around the black hole due to previous missed collisions may affect the gas evolution at later times.\\

Depending on the importance of the effects described above, the subsequent gas evolution can differ significantly. To determine the impact of relativistic precession alone, we start by making the idealized assumption that both the nozzle shock and nodal precession have a negligible impact on the resulting self-crossing shock. The stream components can therefore be approximated as having the same width $H_1 = H_2$ and being perfectly aligned with $\Delta z = 0$. In this situation, the collision takes place in a confined region and involves all the mass of the two streams that intersect with opposite radial velocities $\pm\mathbf{v_{\rm r}}$ as indicated in the left panel (a) of Fig. \ref{fig:sketch}. Because the inflow rate through the collision point is similar\footnote{Due to the time-dependence of the fallback rate, the inflow rate through the collision point may differ for the two stream components, especially if the intersection happens at large distances with $\rint \gtrsim \amin$.} for the two components, the specific energy dissipated by the interaction reaches its maximal value
\begin{equation}
\Delta \varepsilon^{\rm max}_{\rm sc} = \frac{v^2_{ \rm r}}{2} \approx \frac{G \mh}{R_{\rm int}} \sin^2 (\psi/2),
\label{eq:self-crossing}
\end{equation}
since all the kinetic energy associated with radial motion gets converted into heat. In the second step, the radial velocity is expressed as $v_{\rm r} = |\mathbf{v_{\rm r}}| =  v_{\rm int} \sin (\psi/2)$ and the gas velocity at the self-crossing shock is approximated by the local escape speed with $v_{\rm int} \approx (2 G \mh/R_{\rm int})^{1/2}$. Except for very low black hole masses, $\psi \approx \pi$, so that a significant fraction of the binding energy at the intersection radius gets dissipated. A dissipated energy $\Delta \varepsilon_{\rm sc} \approx \Delta \varepsilon_{\rm circ}$ requires $\rint \approx \rp$, which is only attained for $\mh \gtrsim 10^7 \msun$ or $\beta \gtrsim 5$, as explained above. Only in this situation can we expect a fraction of the gas to reach a significant level of circularization as a result of the first collision, although even in this case, a large fraction of the gas may be unbound from the black hole. Otherwise, relativistic precession is weaker, with $\Delta \phi \ll \pi$, often implying $\rint \approx \amin$ and $\Delta \varepsilon_{\rm sc} \approx 10^{-2}  \Delta \varepsilon_{\rm circ}$. Following this interaction away from pericenter, most of the gas remains therefore significantly eccentric. 

\begin{figure}
\centering
\includegraphics[width=\textwidth]{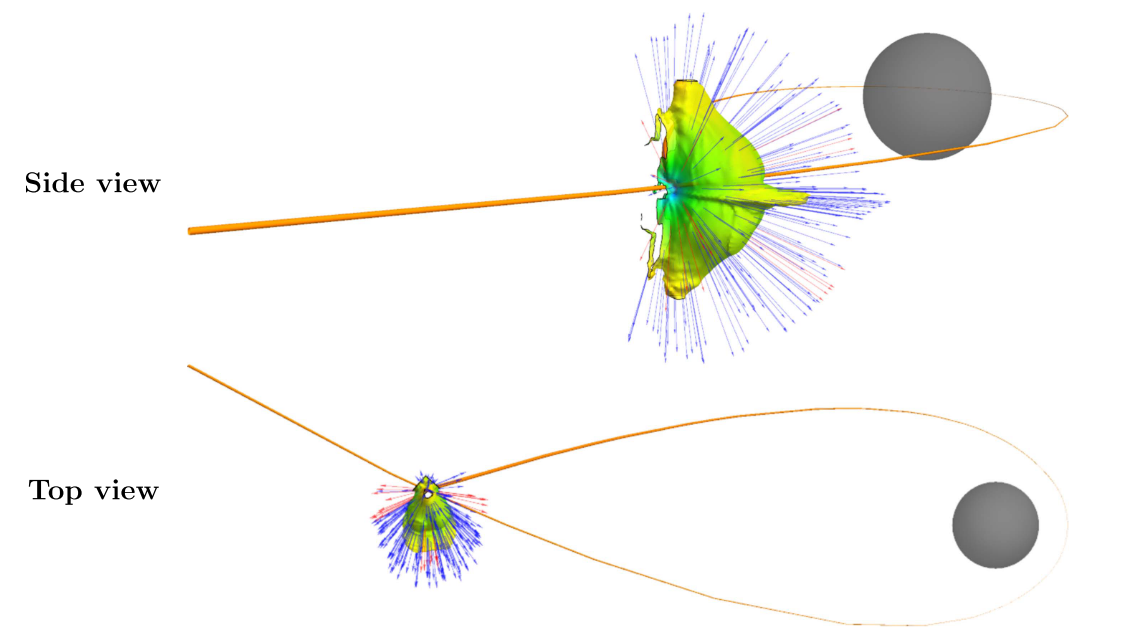}
\caption{Gas evolution resulting from the self-crossing shock as obtained from the local radiative transfer simulation of \cite{jiang2016} with the two panels adopting different points of view. The stream (orange tube) experiences apsidal precession during its passage close to the black hole (grey sphere) that results in its intersection. Due to the pressure increase, the ensuing self-crossing shock causes the formation of an outflow launched from the intersection point displayed with the blue to yellow surface. The arrows represent the gas velocity for the bound (blue) and unbound (red) matter.}
\label{fig:jiang}
\end{figure}

The hydrodynamics of the self-crossing shock has been investigated through local simulations initialized with two identical thin stream components whose properties are obtained analytically \citep{kim1999,jiang2016,lu2020}. As we explain in Section \ref{sec:shock-driven}, the collision is found to be radiatively inefficient in this limit due to the large optical depths of the confined gas such that the evolution can be accurately approximated as adiabatic. These numerical works demonstrate that the shocked gas gets launched into an outflow driven by radiation pressure.\footnote{The self-crossing shock leads to an increase of the gas temperature to $T_{\rm g} \approx m_{\rm p} v^2_{\rm r} / k_{\rm B} \approx  10^{9}\, \rm K$ using equation \eqref{eq:self-crossing} and a typical radial velocity $v_{\rm r} \approx 0.01 \, c$ for the colliding gas. This shocked matter rapidly cools by emitting photons that results in an increase of radiation energy while that of the gas diminishes. At the end of this process, most of the energy is in the form of photons with an equilibrium temperature of $T_{\rm eq} \approx  (\rho_{\rm s}  v^2_{\rm r}/a)^{1/4} \approx  10^6 \, \rm{K}$, where in the lattermost equality we have adopted a density inside the stream of $\rho_{\rm s} \approx 10^{-7} \gpercc$ as derived in Section \ref{sec:stream-evolution} (see Fig. \ref{fig:stream-density}). The ratio of gas to radiation pressure is then $P_{\rm g} / P_{\rm r} \approx (\rho k_{\rm B} T_{\rm eq}/m_{\rm p}) / (a T^4_{\rm eq}/3) \approx 10^{-3}$.} It can be seen from Fig. \ref{fig:jiang} that corresponds to the local radiation-hydrodynamics simulation of a strong intersection by \citet{jiang2016} where the two thin, intersecting stream components are displayed as orange tubes while the outflow is represented with the blue to yellow surface. An idealized version of this outflowing gas is also illustrated by the orange sphere delimited by black dashes in the left panel (a) of Fig. \ref{fig:sketch}. In addition to this quasi-spherical expansion due the excess internal energy given by equation \eqref{eq:self-crossing}, the gas also retains a net velocity with respect to the black hole inherited from the tangential motion of the colliding streams. The velocity vector of this outflowing matter can therefore be approximately decomposed as $\mathbf{v}_{\rm out} = \mathbf{v}_{\rm e} + \mathbf{v}_{\rm n}$ where $|\mathbf{v}_{\rm e}| \approx v_{\rm r}$ and $\mathbf{v_{\rm n}} \approx \mathbf{v_{\rm t}}$, as illustrated in the left panel (a) of Fig. \ref{fig:sketch}. This implies that the gas expanding near the direction of its net motion has a velocity $|\mathbf{v}_{\rm n}|+|\mathbf{v}_{\rm e}| > (\mathbf{v}^2_{\rm r}+\mathbf{v}^2_{\rm t})^{1/2} = v_{\rm int} \approx (2G \mh/\rint)^{1/2}$ that is larger than the local escape speed (purple dotted arrows). The fraction of debris unbound by the collision becomes significant if the self-crossing shock is strong. For $\beta =1$, unbinding more than 20\% of the colliding gas requires a black hole mass $\mh \gtrsim 5 \times 10^6 \msun$ according to local adiabatic simulations \citep{lu2020}.\footnote{For the strong self-crossing shock they consider, the radiation-hydrodynamics simulation performed by \cite{jiang2016} finds that about $16\%$ of the debris gets unbound, which is lower than the value of $50\%$ obtained from the adiabatic study of \cite{lu2020}. A possible origin for this discrepancy is that radiative losses reduce the impact of radiation pressure on the gas that therefore gets accelerated to a lower terminal speed, although it may also result from 
differences in the numerical setup used or the intial properties of the colliding streams.} Similarly, the gas that gets accelerated in the direction opposite to its net speed (green dotted arrows) experiences a change in sign of its angular momentum if $|\mathbf{v}_{\rm n}| < |\mathbf{v}_{\rm e}|$, so that it ends up rotating in the direction opposite to that of the original star around the black hole.

Most of the local studies dedicated to the self-crossing shock have been made in the idealized situation we just discussed.\footnote{Note that the self-crossing shock found in most global simulations of disc formation presented in Section \ref{sec:hydrodynamics} actually differ from the ideal situation of identical stream components, but this difference may, at least partially, originate from the simplified initial conditions used in these studies.} We present now possible consequences of a more realistic configuration involving two stream components that are misaligned with different widths. For similar widths $H_1 \lesssim H_2$ and a mild vertical offset $\Delta z \ll H_1+H_2$, only a small fraction of the colliding streams can avoid the interaction. Additionally, the two stream components may reach the intersection point with slightly different inflow rates. These effects tend to reduce the dissipated specific energy below the maximal value of equation \eqref{eq:self-crossing}, i.e. $\Delta \varepsilon_{\rm sc} \lesssim \Delta \varepsilon^{\rm max}_{\rm sc}$. As a result, the shocked gas will expand slower, with less unbound mass, and may evolve into an outflow that deviates from a quasi-spherical wind due to the asymmetry of the collision. Despite these possible differences, the gas evolution is likely to remain qualitatively similar to that predicted above for identical streams. On the other hand, if the widths of the two stream components are very different, with $H_1 \ll H_2$ due to a fast expansion induced by the nozzle shock, most of the outgoing gas can avoid a direct collision. Due to its much larger density, the infalling debris would be largely unaffected by the interaction and act instead as a static obstacle. As a result, a bow shock could form inside the thicker stream component that may cause a slow expansion of this gas when it continues to move outward past the intersection, as is illustrated with the orange half-sphere in panel (b) of Fig. \ref{fig:sketch}. The energy dissipated during this process is likely much lower than that given by equation \eqref{eq:self-crossing}, with $\Delta \varepsilon_{\rm sc} \ll \Delta \varepsilon^{\rm max}_{\rm sc}$, while the radiative efficiency could be significantly higher than mentioned above, as we discuss more in detail in Section \ref{sec:shock-driven}. If the outgoing stream component has a width $H_2 \approx \rint$ similar to the intersection radius of equation \eqref{eq:intersection_radius}, the self-crossing-shock takes place at a wide range of distances from the black hole due to a deflection of its trajectories induced by the nozzle shock. Despite the overall weak dissipation, a fraction of the debris could collide near pericenter due to a strong effective precession that may result in its circularization. As we mentioned before, this effect could be important for low-mass black holes with $\mh \lesssim 10^5 \msun$. These important differences caused by the increased gas thickness suggest that the hydrodynamics in this regime qualitatively differs from the case of identical streams. However, additional work is needed to better understand the gas evolution during a realistic self-crossing shock.

As we have seen, several possible outcomes are possible for the self-crossing shock depending on how this interaction takes place. It is, however, still largely unclear for which parameters of the problem they are each realised, due to uncertainties in the mechanisms involved, most importantly the expansion at the nozzle shock. Collisions near pericenter are necessary to dissipate a specific energy $\Delta \varepsilon_{\rm sc} \approx \Delta \ecirc$ but they require either strong relativistic apsidal precession or a large expansion induced by the nozzle shock that only seems realised in specific conditions. In this case, the self-crossing shock itself could result in a significant level of circularization. More generally, interactions happen near the apocenter of the most bound debris with $\Delta \varepsilon_{\rm sc} \ll \Delta \ecirc$ such that the shocked gas retains large eccentricities. More dissipation would therefore be necessary for the gas trajectories to further circularize. In each regime, the gas remaining bound following the self-crossing shock comes back towards the black hole at later times, as indicated with the dotted arrows in the two panels of Fig. \ref{fig:sketch}. During its subsequent evolution, this matter is likely to experience more interactions, as will be discussed in Section \ref{sec:hydrodynamics} based on global simulations.

\subsection{Computational challenges and treatment of general-relativistic effects}
\label{sec:numerical}

In Section \ref{sec:initial}, we described the early dynamics followed by the debris based on analytical arguments and local numerical studies of the self-crossing shock that apply for the realistic situation of a star disrupted on a parabolic orbit by a supermassive black hole. Studying the more complex evolution of this matter at later times requires performing a global simulation of the disc formation process. However, this task turns out to be very numerically challenging for the following reasons. The fluid elements inside the stream have semi-major axes larger than that of the most bound debris  $\amin$ while the returning matter evolves on scales similar to $\rp \ll \amin$ that are two to three orders of magnitude smaller. The computational cost associated with this large dynamic range is exacerbated by the high aspect ratio of the stream due to the transverse confinement by self-gravity that limits its width to $H \approx 10 \rstar \ll \rp$ (see Section \ref{sec:basic}). This numerical challenge has  long been recognized. The first hydrodynamical simulation of the disc formation process by \cite{ayal2000} was realised for a realistic setup using a post-Newtonian smoothed-particle hydrodynamics technique. Despite the particle splitting technique used, it is now clear that the numerical resolution of this work is insufficient to accurately follow the gas evolution near pericenter, leading to important numerical artefacts such as the unbinding of most of the returning stream.\footnote{Analytically, this unbinding of mass at pericenter is disfavoured by the fact that the energy dissipated by the nozzle shock is much less than the binding energy of the most bound debris with a ratio $\Delta \varepsilon_{\rm no} / \Delta \varepsilon = \beta^2 (\mh / \mstar)^{-1/3} \approx 0.01 \ll 1$ using equations \eqref{eq:energy-spread} and \eqref{eq:nozzle}. This effect has also not been found in any of the simulations performed at higher resolution, where the stream instead expands around its original trajectory \citep[e.g.][]{guillochon2014-10jh}. Its numerical origin in the low resolution simulations by \citet{ayal2000} likely results from an inaccurate computation of pressure forces at pericenter. Because this matter moves on near-parabolic trajectories, the resulting variations in velocities may cause the artificial unbinding.} The numerical works investigating the global gas evolution during disc formation are included in Table \ref{tab:simulations} for entries with a first letter `G' in the last column corresponding to the method used. The choice of parameters and the technique employed to account for general-relativistic effects are also indicated. In order to have sufficient resolution, most of these works had to adopt simplified initial conditions that are not astrophysically realistic. Two different ``compromises'' have been used that aim at artificially decreasing the dynamic range of the problem to make it computationally tractable. More recently, \citet{bonnerot2020-realistic} used a third strategy that consists in injecting gas into the computational domain to model the outflow resulting from the self-crossing shock. This method allowed them to perform for the first time a simulation of disc formation for realistic parameters of the problem. We start by describing these approaches below along with the astrophysical limitations of each and then focus on the treatment of general relativistic effects. A detailed discussion of different numerical methods can be found in the \simchap{} while the \fmodchap{} describes innovative techniques that could be used in the future.

The first compromise is to simulate the disruption of a star by an intermediate-mass black hole rather than a supermassive one. Since $a_{\rm min} / R_{\rm t} \propto \mh^{1/3}$ (see equations \eqref{eq:tidal-radius} and \eqref{eq:semi-major-axis}), the dynamic range may be reduced by an order of magnitude by considering a black hole mass of $\mh = 10^3 M_\odot$ instead of a more realistic choice of $\mh =10^6 \msun$. The first such simulations \citep{rosswog2008,ramirez-ruiz2009,rosswog2009} were carried out to study tidal disruptions by intermediate-mass black holes for themselves. One important motivation is that white dwarves can only be disrupted outside the event horizon of the black hole for $\mh \lesssim 10^5 \msun$. Later numerical works \citep{guillochon2014-10jh,shiokawa2015} were instead performed with the intention to get insight into the case of a supermassive black hole. However, disruptions by such low-mass black holes differ from the realistic situation in three main ways that must be kept in mind when attempting to extrapolate the results: (i) the dynamical importance of energy dissipation at the nozzle shock is overestimated as shown in Section \ref{sec:nozzle} \citep{guillochon2014-10jh}, (ii) the pericenter distance is less relativistic since $R_{\rm t}/R_{\rm g} \propto \mh^{-2/3}$ that can reduce the apsidal precession angle\footnote{Relativistic apsidal precession can be important for TDEs involving intermediate-mass black holes if $\beta \gg 1$ but this type of deeply-penetrating encounter is uncommon and poses its own computational challenges related to accurately resolving mid-plane compression in the disrupting star and the returning debris streams.} and strength of the self-crossing shock (see Section \ref{sec:self-crossing}) and (iii) the tidal approximation is itself less generally well-satisfied as $R_{\rm t}/R_\star \propto \mh^{1/3}$. Of these three effects, (i) and (ii) may be quite important for the process of disc formation. It is less likely that (iii) has direct relevance for this phase of evolution, although it may result in a more subtle shift in the energy distribution of the debris \citep{Brassart2008}.

The second numerical compromise that has been made is to simulate the disruption by a supermassive black hole of a star on an initially bound rather than parabolic trajectory. The associated decrease in the semi-major axis $\amin$ of the most-bound debris can significantly reduce the numerical cost of the simulation for stellar eccentricities of $e_{\star} \lesssim 0.95$. Stellar disruptions can occur on bound trajectories in some situations involving the previous tidal separation of a binary star \citep{amaro-seoane2012}, a recoiling \citep{stone2012-recoiled} or binary \citep{coughlin2017-binary} supermassive black hole, as described in the \binchap. However, the main motivation for carrying out such simulations is to extrapolate the results to the realistic case of an initial parabolic trajectory. This technique was first used by \cite{hayasaki2013} and later in several other investigations \citep{bonnerot2016-circ,hayasaki2016-spin,sadowski2016}. Although this approach does not suffer from the issues listed above for the first method since it uses the correct black hole mass, it has its own disadvantages: (i) the stream produced by the disruption of a {\it bound} star is artificially truncated and, if $e_{\star}$ is too small, one winds up in the unrealistic situation where the ``head'' of the stream must fruitlessly chase its ``tail'' for several orbits before any major interaction takes place and (ii) the debris stream produced by the disruption process is artificially bound with an apocenter much less than $a_{\rm min}$, which artificially enhances the strength of the self-crossing shock.

The strategy designed by \citet{bonnerot2020-realistic} consists in modelling the outflow from the self-crossing shock by an injection of gas into the computational domain. This method is promising since it can be applied for both a parabolic orbit of the star and a supermassive black hole mass. Nevertheless, the idealized treatment of the self-crossing shock implies the following main caveats: (i) the properties of this interaction are fixed, which neglects any feedback resulting from the formation of the disc such as its impact on the stream trajectory near pericenter and (ii) this strategy requires that the self-crossing shock can be treated as a local interaction, which may not be possible if one of the colliding stream components has a width similar to the intersection radius (see right panel of Fig. \ref{fig:sketch}).

General-relativistic effects are essential to properly describe the gas dynamics during disc formation and must therefore be taken into account to simulate this phase of evolution. In particular, this process is thought to be initiated by a self-crossing shock (see Section \ref{sec:self-crossing}) that depends on both relativistic apsidal and nodal precession, the latter being induced by the black hole spin. Despite the large computational cost of this approach, some numerical works \citep[e.g.][]{shiokawa2015,sadowski2016} adopt a fully general-relativistic treatment. To reduce this numerical expense, other investigations take these effects into account through the use of gravitational potentials that differ from the usual Keplerian one of $\Phi = -G\mh/R$ while still using a Newtonian description of the hydrodynamics.\footnote{Note that several investigations \citep{guillochon2014-10jh,bonnerot2016-circ} nevertheless adopt an entirely Keplerian description. This approach is legitimate when the precession angle is small but can also be useful to identify whether some features of the gas evolution occur independently of general relativistic effects.} The first approach adopted by several authors \citep{rosswog2009,hayasaki2013,bonnerot2016-circ} is to use so called pseudo-Newtonian potentials that are designed by hand to reproduce certain features of general-relativistic gravity. Popular choices include the famous Paczynski-Wiita potential \citep{paczynsky1980} and those designed by \citet{wegg2011} and \citet{tejeda2013}. The latter two are especially relevant since they are tailored to reproduce apsidal precession rates in a Schwarzschild spacetime. A second option is to use post-Newtonian potentials \citep[e.g.][]{blanchet2014}, which are self-consistently derived from the low-velocity limit of general relativity. This is the approach that was used in the first simulations of mass return in TDEs \citep{ayal2000}.  However, the appeal of self-consistency is often outweighed by the poor convergence of the post-Newtonian approximation as distances approach $\rg$. Test particle dynamics in such potentials are increasingly untrustworthy at $R \lesssim 15 R_{\rm g}$, where large deviations from general-relativistic geodesics occur, although it is a good approximation at larger scales. The main advantage of this approach is that it offers a well-defined way to incorporate nodal precession due to black hole spin \citep{faye2006} that has been exploited by \cite{hayasaki2016-spin}.

\subsection{Global hydrodynamics of accretion flow formation}

\label{sec:hydrodynamics}

\begin{table*}
\begin{threeparttable}
    \centering
    \begin{tabular}{c c c c c c c}
    \hline
    Reference & $\mh$ ($\msun$) & a & $e_{\star}$ & $\beta$ & Method \\
    \hline
    \cite{kim1999} & $10^6$ & 0 & 1 & 1 & L/N \\    
    \cite{ayal2000}\tnote{$\star$} & $10^6$ & 0 & 1 & 1 & G/GR \\    
    \citetalias{ramirez-ruiz2009}\tnote{$\dagger$} & $10^3$ & 0 & 1 & 3 & G/N \\
    \cite{rosswog2009}\tnote{$\dagger$} & $100$--$10^4$ & 0 & 1 & 0.9 -- 12 & G/PNP \\
    \cite{haas2012}\tnote{$\dagger$} & $10^3$ & 0--0.6 & 1 & 6--8 & G/GR \\
    \cite{dai2013} & $10^7$ & 0--0.5 & 0.7--0.99 & 1 & B/GR \\    
    \cite{hayasaki2013} & $10^6$ & 0 & 0.8 & 5 & G/PNP \\
    \cite{guillochon2014-10jh} & $10^3$ & 0 & 1 & 1 & G/N \\
    \citetalias{guillochon2015} & $10^5$--$10^8$ & 0--1 & 1 & $\geq 1$ & B/PNC \\    
    \cite{shiokawa2015} & $500$ & 0 & 1 & 1 & G/GR \\
    \cite{evans2015}\tnote{$\dagger$} & $10^5$ & 0 & 1 & 10--15 & G/GR \\
    \cite{dai2015} & $10^5$--$10^8$ & 0 & 1 & 1--16 & B/PNC \\
    \citet{bonnerot2016-circ} & $10^6$ & 0 & 0.8--0.95 & 1--5 &  G/PNP \\
    \cite{hayasaki2016-spin} & $10^6$ & -0.9--0.9 & 0.8 & 1--5 &  G/PNC \\
    \cite{sadowski2016} & $10^5$ & 0 & 0.97 & 10 & G/GR \\
    \cite{jiang2016} & $10^5$--$10^7$ & 0--0.8 & 1 & 0.6--1.2 & L/N \\      
	\cite{bonnerot2017-stream} & $10^5$--$10^7$ & 0 & 1 & 1--5 & B/PNC \\       
	\cite{liptai2019-spin} & $10^6$ & -0.99--0.99 & 0.95 & 1--5 & G/GR \\	
	    \cite{lu2020} & $10^5$--$10^8$ & 0 & 1 & 1--2 & B/GR+L/N \\
	\cite{bonnerot2020-realistic} & $2.5 \times 10^6$ & 0 & 1 & 1 & G/PNP \\

    \end{tabular}
\begin{tablenotes}
\item[$^{\star}$] Although the early work by \citet{ayal2000} carried out a global simulation using realistic parameters of the problem, it is now clear that most of the gas evolution they found results from numerical artefacts associated to the low numerical resolution used.
\item[$^{\dagger}$] These global simulations explicitly aim at studying disruptions by intermediate-mass black holes while the others involving low black hole masses attempt to extrapolate their results to the case of a supermassive one.
\end{tablenotes}    
    
    \caption{Chronological list of the works investigating the gas evolution leading to disc formation in TDEs and the choice of parameters adopted. The method used is specified in the last column by two sets of letters divided by a slash, with the first one corresponding to the numerical strategy used to evolve the gas and the second indicating how the black hole's gravity is treated. The first letter `B' stands for a ballistic evolution as test particles using a simplified analytical treatment for hydrodynamics, `L' for local hydrodynamical simulations that focus on a small domain as convenient to study the self-crossing shock and `G' for global simulations that follow the hydrodynamics for a large fraction of the volume, as necessary to study disc formation. For the second set of letters after the slash, `N' stands for Newtonian gravity, `PNP'  for pseudo-Newtonian potential, `PNC' for post-Newtonian corrections, and `GR' for methods that either fully solve the Einstein's field equations or assume a fixed background metric to evolve the gas.}
    \label{tab:simulations}
\end{threeparttable}
\end{table*}

As explained in Section \ref{sec:numerical}, it is not yet computationally feasible to follow the hydrodynamics of disc formation for realistic initial conditions. The global simulations listed in Table \ref{tab:simulations} therefore rely on computationally-motivated idealizations to make the problem tractable. In most of them, this is achieved by either considering an intermediate-mass black hole or an initially bound star. More recently, \cite{bonnerot2020-realistic} were able to consider realistic astrophysical parameters of the problem using another simplifying strategy: treating the self-crossing shock as an injection of gas into the computational domain. Most of these works adopt an adiabatic equation of state for the gas while a minority of others \citep[e.g.][]{hayasaki2013} assumes that the debris evolves instead isothermally. The adiabatic choice is motivated by the radiative inefficiency of the self-crossing shock if it involves thin stream components (see Section \ref{sec:self-crossing}) but it is unclear whether this assumption remains acceptable in the later stages of disc formation. As we discuss in Section \ref{sec:shock-driven}, it is possible that the circularizing shocks become more radiatively efficient as the accretion flow starts forming, with the hydrodynamics significantly deviating from complete adiabaticity. Most of the numerical works below focus entirely on the formation of the disc and do not follow the subsequent phase of accretion onto the black hole, which is the main topic of the \diskchap{} that we also discuss separately in Section \ref{sec:nascent}. The impact of magnetic fields that are expected to drive accretion is examined in only one of these investigations \citep{sadowski2016} while the others consider a purely hydrodynamical evolution. Accretion driven by magnetized turbulence, as well as additional dissipative processes (e.g. interactions with the stream of debris that continues impacting the accretion disc) may lead to disc evolution that significantly changes the disc properties subsequent to formation. In the following, we describe the ways in which the simulated debris evolves towards an accretion flow, and the properties of the newly-formed discs obtained from these simulations. All these works find that a majority of the bound gas has its average eccentricity decreased due to the interactions it experiences, resulting in a more circular gas distribution. We are particularly interested in the ``circularization timescale,'' $\tcirc$, required for this process to complete, keeping in mind that the resulting accretion flow is in general not entirely circular with the gas retaining instead non-zero eccentricities. Because these simulations use idealized initial conditions, it is not a simple task to extrapolate from these numerical results to the astrophysically realistic situation. We attempt to highlight when this seems nevertheless possible and mention possible differences if such a generalization appears unreliable.

\begin{figure}
\centering
\includegraphics[width=\textwidth]{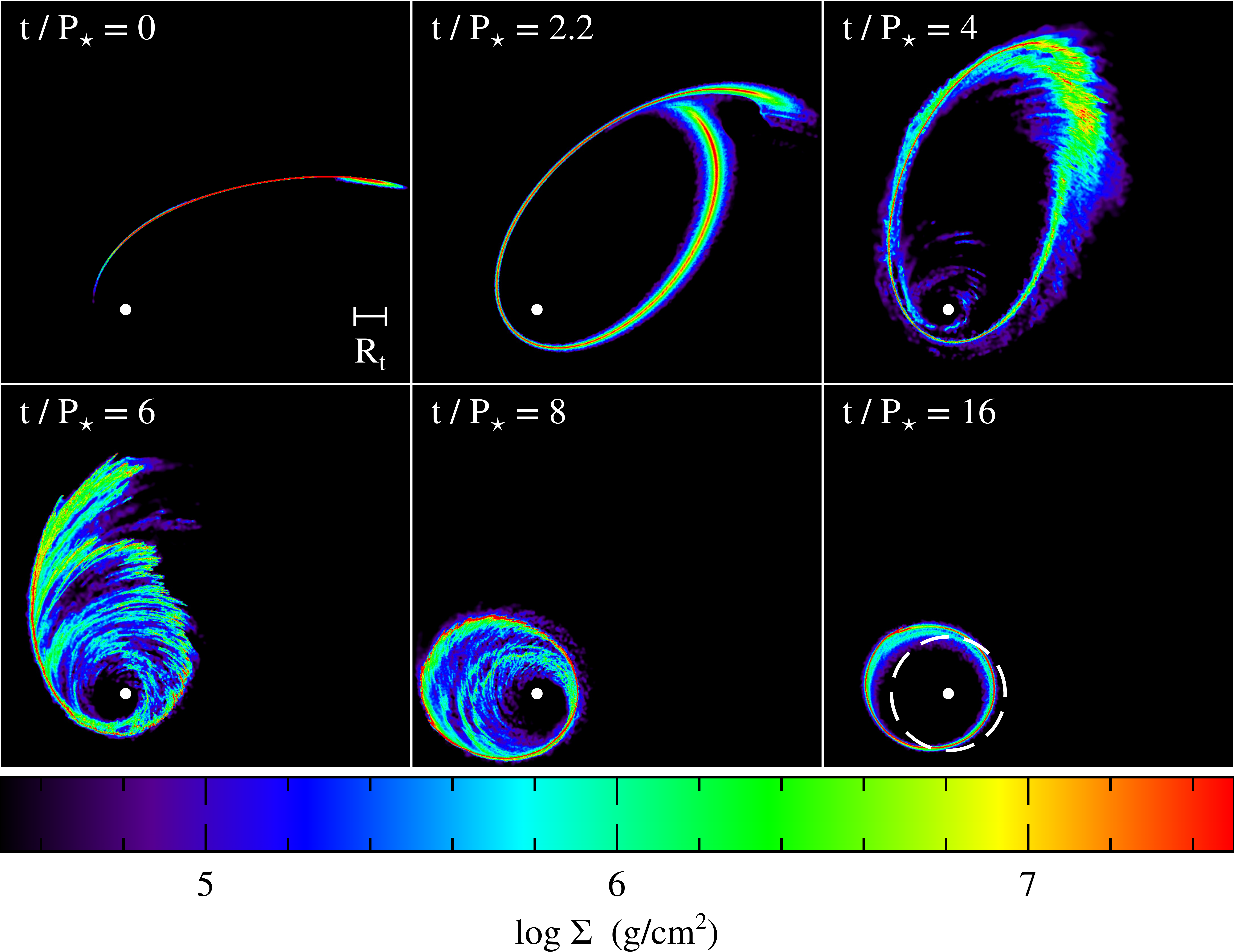}
\caption{Snapshots showing the evolution of column density during the disc formation process simulated by \cite{bonnerot2016-circ}, considering the disruption of a solar-type star on a trajectory with eccentricity $e_{\star} = 0.8$ and penetration factor $\beta=1$ by a supermassive black hole of mass $\mh = 10^6 \msun$. Due to weak relativistic precession, the stream intersects itself near apocenter, and the debris circularizes on a timescale $\tcirc \approx 10 P_{\star}$. Importantly, the gas is assumed to evolve isothermally, resulting in the formation of a thin ring located around the circularization radius $R_{\rm circ} \approx 2 \rp$ (dashed circle in the bottom right snapshot).}
\label{fig:bonnerot-bound}
\end{figure}

\begin{figure}
\centering
\includegraphics[width=\textwidth]{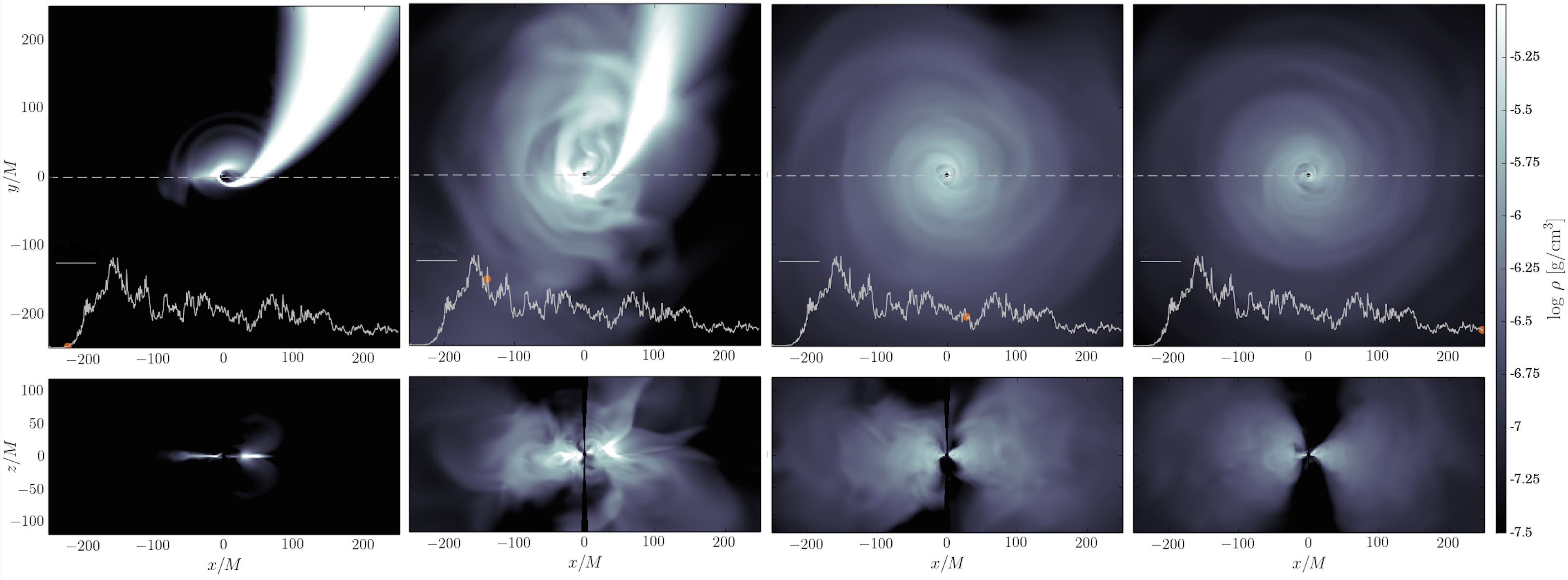}
\caption{Snapshots from the simulation by \cite{sadowski2016}, that studied the disc formation process for a star with eccentricity of $e_{\star} = 0.97$ and penetration factor $\beta = 10$, disrupted by a black hole of mass $\mh = 10^5 \msun$. The density distribution is shown in slices parallel (upper panels) and perpendicular (lower panel) to the stellar orbital plane. Due to strong relativistic precession, a self-crossing shock takes place near pericenter, resulting in a fast outflow and the formation of an accretion flow on a timescale $\tcirc \approx 5 P_{\star} \ll \tmin$. Because the gas is assumed to evolve adiabatically, the resulting disc is extended and thick.}
\label{fig:sadowski}
\end{figure}

The first hydrodynamic simulation following accretion flow formation in its entirety is that by \citet{hayasaki2013}, who consider a bound star with $e_{\star} = 0.8$ and $\beta = 5$ disrupted by a supermassive black hole of mass $\mh =10^6 \msun$. Due to the low stellar eccentricity, all the debris is bound, unlike the situation in a parabolic encounter. As explained in Section \ref{sec:numerical}, this causes the stream to chase its ``tail'' for a few orbits before it first intersects itself due to strong relativistic apsidal precession. Collisions near pericenter then happen at each revolution, although the first ones only involve the stream's low-density extremities. As a result, an accretion disc forms on a timescale $\tcirc \approx 5 P_{\star}$ where $P_{\star}$ denotes the stellar period.\footnote{Because of the bound stellar trajectory, the difference in period between the different parts of the stream is negligible compared to the period of the star. For this reason, when discussing idealized simulations with significantly sub-parabolic stellar eccentricities, we express the circularization timescale in terms of the stellar period rather than that of the most bound debris.} Later simulations \citep{bonnerot2016-circ,hayasaki2016-spin} explored the impact of the stellar trajectory and gas thermodynamics on this process. They find that an increase in eccentricity to $e_{\star} \approx 0.95$ allows the stream to promptly cross itself for $\beta=5$ without the prior revolutions artificially needed for a more bound star. Due to the large precession angle, the ensuing self-crossing shock is strong, causing rapid formation of an accretion disc on a reduced timescale of $\tcirc \approx P_{\star}$. However, decreasing the penetration factor to a more common value of $\beta = 1$ results in collisions near the apocenter of the stream due to the decreased relativistic precession angle, as described in Section \ref{sec:self-crossing}. As can be seen from Fig. \ref{fig:bonnerot-bound}, these weak interactions result in a slower disc formation with $\tcirc \approx 10 P_{\star}$. As in the earlier work by \citet{hayasaki2013}, this evolution is somewhat delayed by the bound stellar trajectory considered. The gas also does not significantly expand during the circularizing shocks, causing the formation of a thin circular ring at $R_{\rm circ} \approx 2 \rp$. Importantly, this outcome directly results from the assumed isothermal equation of state, that allows the gas to effectively lose energy at shocks. As emphasized above, this assumption is unlikely to be valid in most cases due to the large optical thickness of the colliding streams. By modifying the equation of state to the more physical adiabatic limit, \citet{hayasaki2016-spin} and \citet{bonnerot2016-circ} find that the stream expands as a result of collisions. However, this additional expansion does not unbind any gas from the black hole, which is likely due to the artificially bound stellar orbit considered. In this case, the resulting accretion disc is thick and occupies a wide range of radii centred around the stellar semi-major axis. 

\citet{sadowski2016} also studied disc formation assuming adiabaticity for an eccentric stellar trajectory, with $e_{\star}=0.97$ and a large penetration factor of $\beta =10$. As can be seen from the snapshots of Fig. \ref{fig:sadowski}, strong relativistic precession then results in a self-crossing shock near pericenter. This collision involves stream components of different widths with $H_2 \gtrsim H_1$ (see Fig. \ref{fig:sketch}) due to a fast expansion caused by the nozzle shock during pericenter passage, as discussed in Section \ref{sec:nozzle}. Despite this asymmetry, the collision is powerful enough to cause the shocked gas to expand into an outflow whose unbound component represents about $15 \%$ of the stellar matter. While this unbinding of matter is consistent with the expectation of Section \ref{sec:self-crossing}, the gas retains its original direction of rotation around the black hole, in contrast to the change of angular momentum sign predicted by local studies for a strong self-crossing shock. The bound part of this post-shock gas then returns near the black hole with a significant fraction being ballistically accreted along the funnels of the forming disc. After experiencing numerous interactions, the debris settles on a timescale $\tcirc \approx 5 P_{\star}$ into an extended and thick accretion flow. As in the prior works that assume adiabaticity, this disc extends roughly to the stellar semi-major axis, with significant pressure support against gravity in its outer region. Despite its overall axisymmetric structure, the gas retains moderate eccentricities with $e\approx 0.2$ on average. This work was also the first to include gas magnetic fields, but these were found not to modify the hydrodynamics, as we discuss more in detail in Section \ref{sec:viscosity}. In this study as well as the above simulations with $e \gtrsim 0.95$ and $\beta \gtrsim 5$ \citep[e.g.][]{bonnerot2016-circ} that are able to capture a prompt self-crossing shock, the gas evolution may be comparable to that of a deep parabolic encounter, although the precise properties of the initial collision could significantly differ (see Section \ref{sec:self-crossing}). Because the reduced period of the stream is washed out by the large redistribution of orbital elements induced by multiple collisions, it appears legitimate to compare the circularization timescale of a few $P_{\star}$ found in this regime to the period $\tmin$ of the most bound debris assuming a parabolic stellar trajectory. This comparison yields $\tcirc \lesssim \tmin$, which suggests that disc formation happens rapidly for deep encounters where a strong self-crossing shock takes place.\footnote{A more extreme situation is that of an ultra-deep encounter where the stellar pericenter is similar to the gravitational radius of the black hole. In this case, the disruption of the star is accompanied by a stretching of the debris into a elongated structure. Due to the very large values of the apsidal precession angle with $\Delta \phi \gtrsim \pi$, it is possible that this gas collides with itself during the first passage of the star near pericenter. Numerical investigations of this process \citep{haas2012,evans2015,darbha2019} find that this early self-crossing shock results in the fast formation of an accretion flow around the black hole. We note however that such relativistic pericenters are rare for black holes substantially smaller than the Hills mass.} For $\beta \approx 1$, extrapolating eccentric disruption results to the parabolic case is more uncertain due to the reduced size of an entirely bound stream, which can have a larger influence on the gas evolution.

\begin{figure}
\centering
\includegraphics[width=\textwidth]{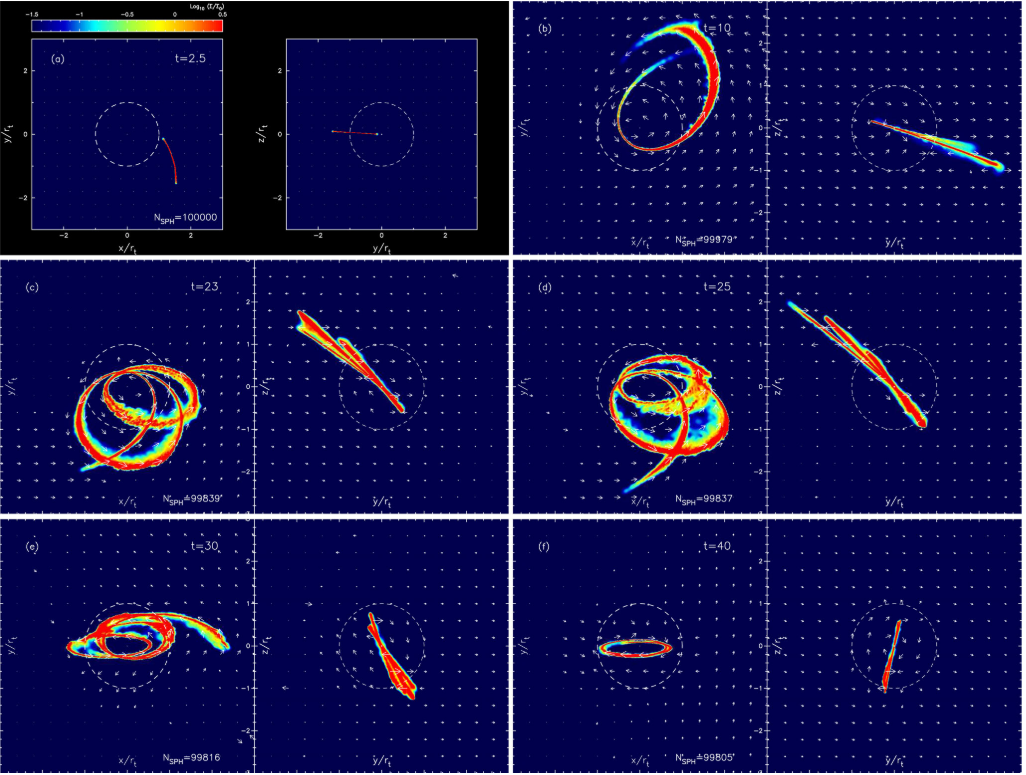}
\caption{Snapshots from the simulation by \cite{hayasaki2016-spin} who study the disc formation process for a star with initial eccentricity of $e_{\star} = 0.7$ disrupted by a black hole with spin $a=0.9$ inclined with respect to the stellar orbital plane by an angle $i=90^{\circ}$. Because the gas is assumed to evolve isothermally, it does not expand significantly during pericenter passage such that nodal precession prevents a successful self-crossing shock for several orbits. Interactions eventually take place that cause the delayed formation of a narrow accretion disc on a timescale  $\tcirc \gtrsim 10 P_{\star}$.}
\label{fig:hayasaki}
\end{figure}

\begin{figure}
\centering
\includegraphics[width=\textwidth]{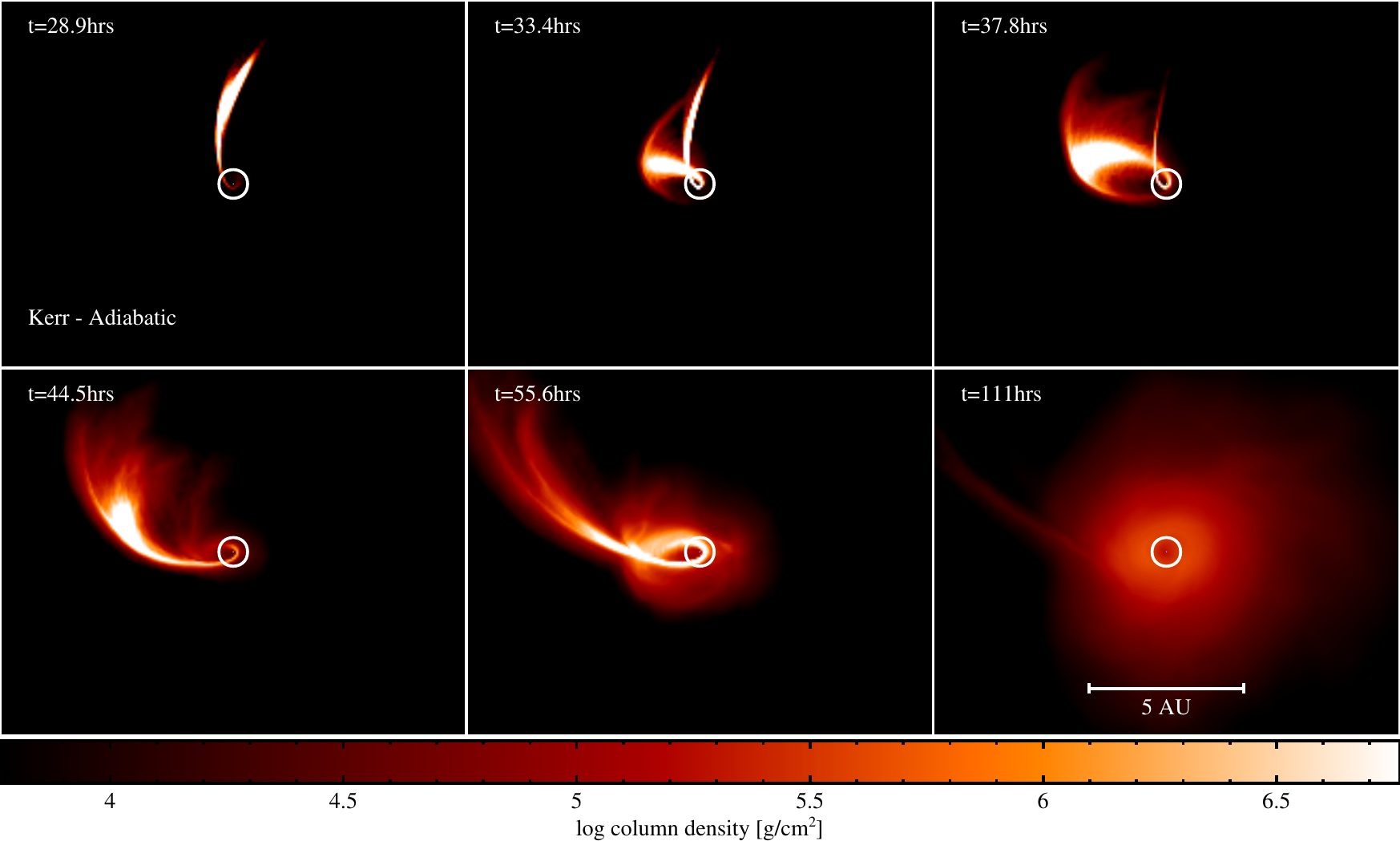}
\caption{Snapshots from the simulation of \cite{liptai2019-spin}, that studies the disc formation process for a star with eccentricity $e_{\star} = 0.95$ and penetration factor $\beta = 5$ disrupted by a black hole with spin $a=0.99$ inclined with respect to the stellar orbital plane by an angle $i=60^{\circ}$. Nodal precession produces a vertical offset that prevents the stream from colliding with itself promptly after its first pericenter passage. However, due to the adiabatic equation of state assumed for the gas, the nozzle shock results in a significant expansion that causes a successful self-crossing shock during the next orbit. This interaction results in the formation of a thick and extended accretion disc that is only mildly delayed, with a circularization timescale $\tcirc \lesssim 5 P_{\star}$.}
\label{fig:liptai}
\end{figure}

The influence of black hole spin has also been studied in simulations considering initially bound stars and different equations of state \citep{hayasaki2016-spin,liptai2019-spin}. When the gas is assumed to evolve isothermally, the energy dissipated by the nozzle shock is removed, so that there is no significant expansion of the stream during pericenter passage. In this case, nodal precession causes a vertical offset $\Delta z > H_1+H_2$ (see Section \ref{sec:self-crossing}) that prevents the stream from colliding with itself even after multiple orbits. This effect is illustrated in Fig. \ref{fig:hayasaki} that shows the gas evolution obtained by \cite{hayasaki2016-spin} for a star with $e_{\star} \approx 0.7$ and $\beta = 2$, disrupted by a black hole with spin $a=0.9$, and inclined by an angle $i=90^{\circ}$ with respect to the orbital plane. A ``wicker basket'' configuration is created as the stream wraps around the black hole until a self-crossing shock eventually takes place causing disc formation to complete after tens of dynamical times with $\tcirc \gtrsim 10 P_{\star}$. These works also find that the nozzle shock makes the stream expand faster when adiabaticity is assumed, which is likely a more physical choice given the large optical depth of the confined gas. In this limit, the delay of the first collision is reduced due to the increase in stream width that can rapidly overcome the vertical offset induced by nodal precession, i.e. $\Delta z < H_1+H_2$. As a result, no delay to the self-crossing shock is observed in the simulations by \citet{hayasaki2016-spin} for the parameters they consider. The more recent investigation by \cite{liptai2019-spin} finds a similar gas evolution, shown in Fig. \ref{fig:liptai} for a stellar trajectory with $e_{\star} = 0.95$ and $\beta = 5$, a black hole spin $a=0.99$, and an inclination angle $i=60^{\circ}$. It can be seen that nodal precession prevents the stream from colliding with itself shortly after its first pericenter passage. However, the fast gas expansion caused by the nozzle shock results in a self-crossing shock during the second orbit. Later on, this interaction induces the formation of a thick accretion disc, that is only mildly delayed with $\tcirc \lesssim 5 P_{\star}$. This conclusion appears qualitatively different from that obtained by \citet{guillochon2015} in the semi-analytical work discussed in Section \ref{sec:self-crossing}, which found that the first collision usually takes place after a large delay of multiple orbital timescales, even when stream expansion near pericenter is taken into account. The discrepancy could be due to an artificially small amount of stream expansion in the analytic model for the nozzle shock compared to that found in simulations\footnote{This difference could originate from the non-zero black hole spin, that may significantly increase the energy dissipated during the nozzle shock if nodal precession causes the formation of oblique collisions, as proposed in Section \ref{sec:nozzle}.} or alternatively to the unrealistically low eccentricities of the debris resulting from the bound stellar trajectory.

\begin{figure}
\centering
\includegraphics[width=\textwidth]{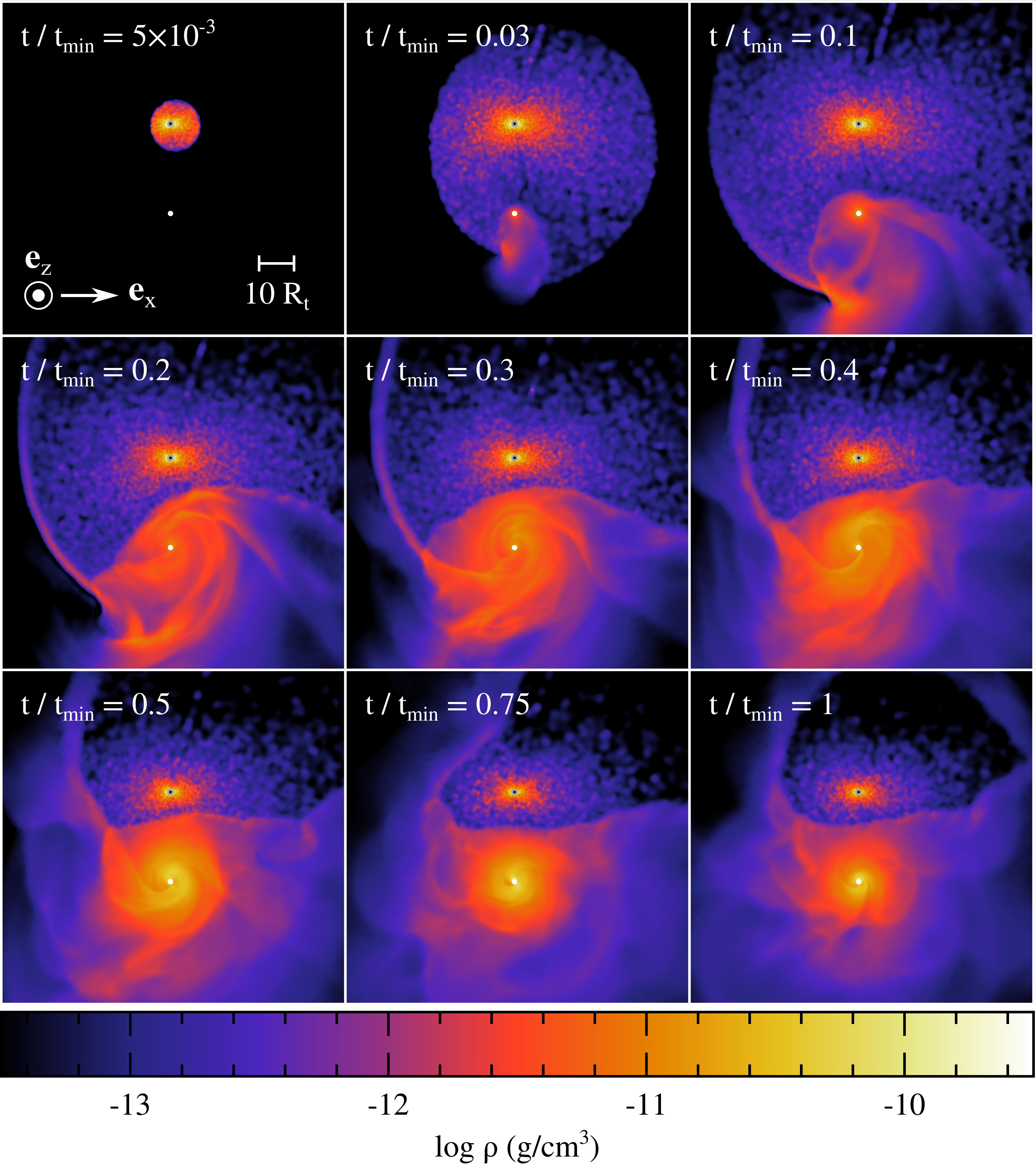}
\caption{Snapshots showing the gas evolution in a slice parallel to the orbital plane of the star during the disc formation process simulated by \cite{bonnerot2020-realistic} for a parabolic stellar trajectory and a supermassive black hole of mass $\mh = 2.5 \times 10^6 \msun$. This numerical work uses as initial condition the outflow launched from a strong self-crossing shock whose properties are obtained from a local study of the collision \citep{lu2020}. Gas is injected inside the computational domain from the intersection point (grey circle) located at $\rint \approx 25 \rt$. This outflowing matter rapidly engulfs the black hole and experiences multiple interactions that result in the formation of a thick and extended accretion disc on a timescale $\tcirc \approx 0.3 \tmin$.}
\label{fig:bonnerot}
\end{figure}

The recent simulation by \cite{bonnerot2020-realistic} uses the outflow launched from the self-crossing shock as an initial condition by artificially injecting gas inside the computational domain from the intersection point. Using this numerical strategy, this work is able to consider realistic parameters of the problem: a parabolic stellar trajectory and a supermassive black hole of mass $\mh = 2.5 \times 10^6 \msun$. The properties of the injected matter are obtained from a local numerical study of the collision \citep{lu2020} assuming that the two stream components have identical and thin widths with $H_1 = H_2 \ll \rint$ (see left panel (a) of Fig. \ref{fig:sketch}). For their choice of parameters, the stream crosses itself relatively close to pericenter with an intersection radius of $\rint \approx 25 \rp$. The ensuing self-crossing shock is strong, leading to the unbinding of about $33 \%$ of the initially bound gas via the mechanism described in Section \ref{sec:self-crossing}, while most of the remaining bound matter has a sign of angular momentum opposite to that of the original star. 
This gas quickly expands into an envelope that completely engulfs the black hole, as can be seen from the snapshots of Fig. \ref{fig:bonnerot} obtained from this simulation. As its trajectories intersect, multiple secondary shocks take place that cause the rapid formation of an accretion disc on a timescale of $\tcirc \approx 0.3 \, \tmin$.  The resulting accretion disc is thick, with significant pressure support due to the adiabatic equation of state used, and retains an average eccentricity of $e \approx 0.2$. The disc is located within a radius $R_{\rm d} \approx 15 \rp$ that contains only a small fraction of the injected gas. Due to the angular momentum sign of the injected bound debris, the accretion flow produced from this matter also rotates in the retrograde direction compared to the star. Additionally, it features two spiral shocks that drive a slow inflow along the disc mid-plane, although most of the accretion onto the black hole is due to gas falling along the funnels of the disc as it assembles. 
Despite the realistic astrophysical parameters it considers, this simulation remains simplified due to its idealized treatment of the self-crossing shock by an injection of gas, as explained in Section \ref{sec:numerical}. In particular, it is unclear whether this collision actually results in the large outflow used as initial conditions, which closely relates to uncertainties mentioned in Section \ref{sec:self-crossing} regarding this early source of dissipation. Nevertheless, the similarity of this work with those presented above that self-consistently simulate a strong self-crossing \citep[e.g.][]{sadowski2016} seems to suggest that the gas evolution is qualitatively correct in this regime.

\begin{figure}
\centering
\includegraphics[width=\textwidth]{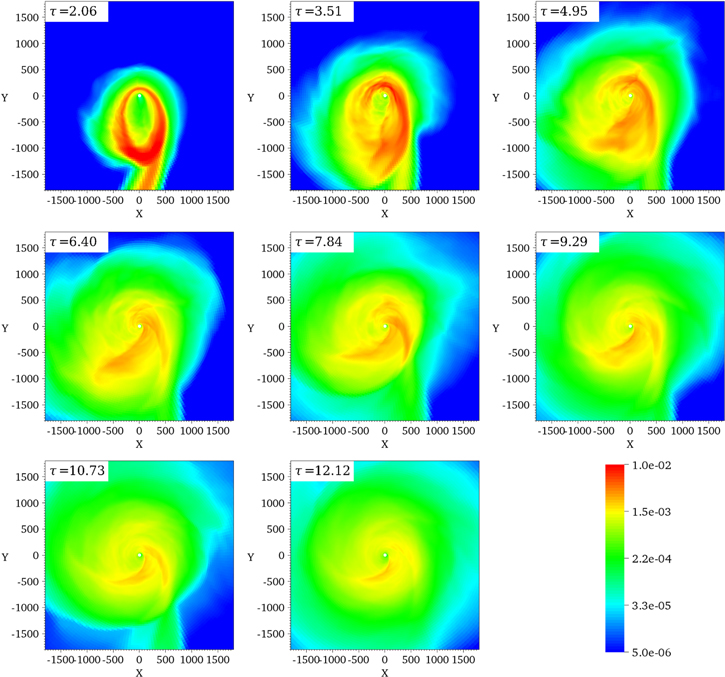}
\caption{Snapshots showing the column density evolution during the disc formation process in the simulation carried out by \cite{shiokawa2015} that consider the disruption of a white dwarf on a parabolic orbit by an intermediate-mass black hole of mass $\mh = 500 \msun$. In the first snapshot, the tip of the stream has already passed pericenter and intersects with the matter still moving inward that results in a self-crossing shock near apocenter. This weak collision initiates the formation of an accretion flow that remains globally eccentric for several orbital periods before eventually settling on a timescale of $\tcirc \approx 10 \, \tmin$ into a more axisymmetric, thick and extended structure.}
\label{fig:shiokawa}
\end{figure}

The simulation by \cite{shiokawa2015} considers the encounter between a white dwarf and an intermediate-mass black hole\footnote{Numerical studies considering intermediate-mass black holes were carried out earlier as well \citep{rosswog2008,rosswog2009,ramirez-ruiz2009,guillochon2014-10jh}. However, they do not run for long enough to capture the completion of disc formation, and we therefore do not present them in detail here.} with $\mh = 500 \msun$ assuming an adiabatic evolution for the gas. Due to weak relativistic apsidal precession, this work finds that the stream intersects itself near the apocenter of its most bound part. This intersection radius is qualitatively similar to that in a large fraction of real TDEs, namely those with $\beta \approx 1$ and $\mh \lesssim 2 \times 10^6 \msun$ (see Section \ref{sec:self-crossing}).\footnote{For the parameters used by \cite{shiokawa2015}, the relativistic apsidal angle is the same as for the disruption of a solar-type star by a black hole of mass $\mh = 3 \times 10^5\msun$ for $\beta =1$.} Around this location, the two colliding components are both thick, with widths $H_1 \approx H_2 \lesssim \rint = \amin$ (see Fig. \ref{fig:sketch}) similar to the intersection radius. This is likely a consequence of the increased dynamical impact of the nozzle shock, which for low-mass black holes causes the stream to significantly expand before it first intersects itself (see Section \ref{sec:numerical}). Interactions between these thick streams are therefore enhanced relative to the weaker collision that might be expected for thin streams, in a way that may compensate for the weak relativistic apsidal precession, as discussed in Section \ref{sec:self-crossing}. As can be seen from the snapshots of Fig. \ref{fig:shiokawa}, the resulting self-crossing shock initially induces mild changes in the gas trajectories associated with a slow expansion that is unable to unbind mass from the black hole. Early on, most of the dissipation still occurs at the nozzle shock, although this interaction weakens due to an expansion of the returning stream. The interactions resulting from the self-crossing shock (termed ``outer shocks'' in this work) are analysed in detail, and decomposed into a ``forward'' and ``reverse'' component that are initially close in space.  At later times, the reverse one recedes closer to the black hole as mass accumulates near apocenter while the forward component remains around the same location. Several passages of the returning gas through this shock system progressively lead to a redistribution of the gas orbital elements accompanied by an overall decrease of its eccentricities. During this process, a significant fraction of the gas has its angular momentum diminished so that it reaches scales similar to the black hole event horizon that can dominate the early accretion, as we discuss more in detail in Section \ref{sec:ballistic}. The accretion flow remains globally eccentric for about ten dynamical times before settling into a more axisymmetric structure (i.e. $\tcirc \approx 10 \, \tmin$) that is thick with internal eccentricities of $e\approx 0.3$ on average, and extends out to a radius $R_{\rm d} \approx \amin$ similar to the semi-major axis of the most bound debris. As was also the case for the weak self-crossing shock captured by the above works \citep[e.g.][]{bonnerot2016-circ} considering bound stars with $\beta \approx 1$, this simulation suggests that circularization is slowed when relativistic apsidal precession is reduced. However, it is unclear whether these results obtained for an intermediate-mass black hole can be extrapolated to the more common situation of a solar-type star disrupted by a supermassive black hole. An important source of uncertainty relates to the increased thickness of the debris, that may significantly modify the gas evolution compared to that involving a thin returning stream confined by self-gravity (see Section \ref{sec:stream-evolution}).

The above simulations aim at studying the global hydrodynamics of accretion flow formation in TDEs. Along with the works presented in Section \ref{sec:initial}, they constitute our current understanding of this complex process, that is to date still highly uncertain, although clearly different from the early prediction by \cite{rees1988}. Despite the simplified initial conditions used, these numerical investigations allow us to gain insight into the astrophysical problem and identify the main remaining sources of uncertainty. If a powerful self-crossing shock takes place near pericenter that strongly modifies the trajectories of the colliding gas, there is convincing evidence that at least a fraction of the debris can promptly form an accretion flow with $\tcirc \lesssim \tmin$, whose properties may be qualitatively similar to those found in some simulations \citep[e.g.][]{sadowski2016,bonnerot2020-realistic}. However, the exact region of parameter space corresponding to this regime and the detailed gas evolution here are still not clearly identified. If the stream intersects itself around apocenter due to weak apsidal precession, some works above \citep{shiokawa2015,bonnerot2016-circ} suggest a much slower formation of the accretion flow. However, it remains to be understood how sensitive these results are to parameter choices made for computational tractability. Tidal disruption by intermediate-mass black holes likely produces streams of much larger relative width due to the enhanced importance of the nozzle shock, while eccentric stellar orbits can lead to unphysical initial interactions between streams (e.g. ``head chasing tail'' behavior) that are probably of greater importance for very weak stream intersections. For these reasons, important uncertainties still exist in this regime, with the possibility that the gas evolves in a realistic situation very differently from what current simulations predict. For example, it is possible that the debris forms a disc that fails to circularize by retaining large eccentricities $e \approx \emin$ for multiple orbital timescales.\footnote{Note that even though the disc could form as very eccentric, it is unclear whether this configuration can be retained due to the various dissipation processes impacting the gas evolution, such as those associated with accretion and the interaction with the stream of loosely bound debris that continues returning to interact with the disc.} Overall, accretion flow formation remains a very open problem, in which the hydrodynamics is not yet qualitatively understood for a large fraction of the parameter space. Progress is likely to come from a better understanding of the early sources of dissipation discussed in Section \ref{sec:initial} combined with improved global simulations more directly applicable to the astrophysically realistic situation.

Regarding the inclusion of additional physics, the influence of black hole spin on this process is also uncertain since it has so so far only been studied for bound stars \citep{hayasaki2016-spin, liptai2019-spin}. The impact of nodal precession at preventing a prompt self-crossing shock remains to be evaluated for a parabolic encounter that would in particular require to estimate the expansion induced by the nozzle shock in this situation (see Section \ref{sec:nozzle}). Another important effect relates to the radiative efficiency, that may increase as the gas distribution becomes more extended due to a decrease of its optical depth, as is explained in more detail in Section \ref{sec:shock-driven}. The gas evolution would then deviate from that assuming complete adiabaticity, which may in particular cause the newly-formed disc to get thinner and more compact as predicted from isothermal runs \citep{hayasaki2013,bonnerot2016-circ, hayasaki2016-spin}. Evaluating the exact level of radiative cooling requires to carry out radiation-hydrodynamics simulations of this process that has not been done so far. In order to capture early gas accretion, it may also be important to include magnetic fields, which has so far been considered in just one simulation \citep{sadowski2016}. Despite the several uncertainties we mentioned, the results of current simulations have led to major improvements in our understanding of accretion flow formation. In the future, additional progress will come from greater computational resources and the development of innovative numerical schemes that would allow us to circumvent the computational limitations presented in Section \ref{sec:numerical}. Once disc formation has been precisely understood, it will become possible to study its longer-term evolution during which most of the gas accretion takes place based on better motivated initial conditions.

\subsection{Shock radiative efficiency and resulting emission}

\label{sec:shock-driven}

Before settling into an accretion disc, the numerical works presented in Section \ref{sec:hydrodynamics} show that the returning debris experiences a large amount of dissipation. The outcome of these shocks depends on their radiative efficiency, that is in turn determined by the optical depth of the surrounding matter. For a very optically thick gas, the photons are advected with the fluid element that created them. In this case, the interaction is radiatively inefficient and the hydrodynamics is close to completely adiabatic. In contrast, more optically thin matter allows photons to be transported away from their production site by diffusion, and to potentially leave the system entirely. The gas evolution may then significantly deviate from the adiabatic limit, and the escaping radiation can participate in the electromagnetic signal observed from TDEs, as discussed more in details in the \emischap{}. We start by evaluating the radiative efficiency of the initial self-crossing shock and then focus on the later interactions that result in accretion flow formation.

As described in Section \ref{sec:self-crossing}, the outcome of the self-crossing shock remains unclear since it depends on several effects whose relative importance has so far not been precisely estimated. As before, we treat first the idealized situation where the collision involves aligned stream components with the same thin width $H_1 = H_2 \ll \rint$, and then discuss possible deviations from these strong assumptions. In this regime, the shock-heating rate is obtained from $\dot{E}^{\rm max}_{\rm sc} = \dot{M}_{\rm fb} \Delta \varepsilon^{\rm max}_{\rm sc}$ by multiplying the fallback rate at which the gas enters the collision by the specific energy dissipated. Using equation \eqref{eq:self-crossing} then yields
\begin{equation}
\dot{E}^{\rm max}_{\rm sc} \approx  \frac{G \mh \dot{M}_{\rm fb}}{\rint} \approx 7 \times 10^{43} \ergpers \left( \frac{\rint}{\amin} \right)^{-1} \left( \frac{\mh}{10^6 \msun} \right)^{-1/6} ,
\label{eq:shock-heating}
\end{equation}
where the numerical value assumes a collision angle $\psi \approx \pi$ and an intersection radius $\rint \approx \amin$ that is valid for the typical parameters $\mh = 10^6 \msun$ and $\beta =1$, and we have set the fallback rate equal to its peak value, with $\dot{M}_{\rm fb} = \dot{M}_{\rm p}$ using equation \eqref{eq:fallback-rate}. This substantial rate of energy dissipation is comparable to the peak optical/UV luminosities observed in typical TDEs, which led  \citet{piran2015-circ} to first propose it as the primary power source for these flares.\footnote{Earlier on, \citet{lodato2012} also estimated the energy that must be radiated for the gas to completely circularize, proposing that this could be an observable power source without further development.} Photons are emitted by the shocked gas with the resulting radiation pressure driving the formation of an outflow that, for simplicity, we approximate as quasi-spherical (see left panel of Fig. \ref{fig:sketch}) with a density profile $\rho_{\rm out} \approx \mdotfb/(4 \pi R^2 v_{\rm out})$ where $R$ denotes the distance from the intersection point. The diffusion and dynamical timescales given by $t_{\rm diff} \approx R \tau_{\rm s}/c$ and $t_{\rm dyn} = R/v_{\rm out}$ are equal at the trapping radius $R_{\rm tr}$ where the scattering optical depth is $\tau_{\rm s} \equiv \int_{R_{\rm tr}}^{\infty} \kappa_{\rm s} \rho_{\rm out} {\rm d} R = c/v_{\rm out}$, yielding $R_{\rm tr} = \mdotfb \kappa_{\rm s}/(4 \pi c)$, where $\kappa_{\rm s} = 0.34 \csperg$ is the electron-scattering opacity. Radiation is coupled to the gas inside this radius and its energy is therefore advected outward at a rate $\dot{E} = 4 \pi e_{\rm r} v_{\rm out} R^2 \propto R^{-2/3}$ where the last scaling uses the fact that the radiation energy density evolves adiabatically as $e_{\rm r} \propto \rho_{\rm out}^{4/3} \propto R^{-8/3}$. This decreasing rate implies that photons lose energy as they are transported outward due to work done on the gas through radiation pressure. After being injected by shocks near the common width $H$ of the two colliding streams, at the luminosity given by equation \eqref{eq:shock-heating}, radiation energy is degraded in this way until photons reach the trapping radius. The emerging luminosity $L^{\rm max}_{\rm sc}$ is therefore reduced by a factor
\begin{equation}
\frac{L^{\rm max}_{\rm sc}}{\dot{E}^{\rm max}_{\rm sc}} \approx \left( \frac{H}{R_{\rm tr}} \right)^{2/3} \approx 0.03 \left(\frac{\mh}{10^6 \msun} \right)^{1/3} \left( \frac{H}{10 \rsun} \right)^{2/3},
\label{eq:shock-luminosity}
\end{equation}
where the numerical value assumes again $\dot{M}_{\rm fb} = \dot{M}_{\rm p}$ and uses a stream width $H \approx 10 \rsun$ that corresponds to an infalling stream confined by self-gravity (see Section \ref{sec:stream-evolution}) \citep{lu2020}. In this regime, the self-crossing shock is therefore radiatively inefficient, a result in agreement with
radiation-hydrodynamics simulations of the collision assuming thin streams \citep{jiang2016}, and that justifies an assumption of adiabaticity for the gas evolution. For $\mh \approx 10^6 \msun$, the resulting luminosity is $L^{\rm max}_{\rm sc} \approx 3 \times 10^{42} \ergpers$, which seems too low to account for the brightest optical and UV emission detected from TDEs. If a significant fraction of the gas collides near pericenter with $\rint \approx \rp$, as is common for $\mh \gtrsim 10^7 \msun$, the heating rate of equation \eqref{eq:shock-heating} could increase and result in a larger luminosity of $L^{\rm max}_{\rm sc} \gtrsim 10^{43} \ergpers$. As discussed in Section \ref{sec:self-crossing}, the two stream components involved in the self-crossing shock may have significantly different widths, with $H_2>H_1$. In this case, we have seen that the specific energy dissipated is likely lower than that used in equation \eqref{eq:shock-heating}, and this would tend to decrease the emerging luminosity compared to $L^{\rm max}_{\rm sc}$. However, a thicker gas distribution accompanied by a significant deviation from spherical geometry for the shocked gas could instead result in a more radiatively efficient interaction, with lower adiabatic losses than is predicted in the above case of identical streams. Our current understanding of the hydrodynamics of a realistic self-crossing shock would have to be improved to go beyond these qualitative expectations.

Following the self-crossing shock, the gas undergoes additional interactions that may circularize the orbits to form an accretion flow. We now focus on the radiation emerging during this later stage, bearing in mind that it may not always be possible to distinguish it from the initial collision, especially if a significant fraction of the gas starts accumulating near the black hole as a result of the self-crossing shock alone.\footnote{This situation is for example expected if early interactions take place near pericenter either as a result of strong relativistic apsidal precession or due to gas deflection induced by a fast expansion at the nozzle shock (see Section \ref{sec:self-crossing}).} If these circularizing shocks can happen closer to the black hole, they are likely stronger than the initial collision. In the work by \cite{bonnerot2020-realistic} that adopts realistic astrophysical parameters, the corresponding heating rate reaches $\dot{E}_{\rm sh} \approx 10^{44} \ergpers$. Notably, \citet{piran2015-circ} extrapolates a similar value from the simulations by \cite{shiokawa2015} that considers an intermediate-mass black hole. Estimating the fraction of radiation that emerges from the system requires to know the gas distribution that the corresponding photons have to go through to escape. This can be done by assuming that this matter has expanded to fill  an approximately spherical region within a distance $R_{\rm d}$ from the black hole. If it contains about the mass of the original star, the optical depth of this gas may be estimated as $\tau_{\rm s} \approx \kappa_{\rm s} M_{\star}/(4 \pi R^2_{\rm d})$. Comparing the diffusion timescale $t_{\rm diff} \approx R_{\rm d} \tau_{\rm s}/c$ to the dynamical time $t_{\rm dyn} \approx 2 \pi (G \mh / R^3_{\rm d})^{-1/2}$ then yields 
\begin{equation}
\frac{t_{\rm diff}}{t_{\rm dyn}} \approx 1.5 \left( \frac{\mh}{10^6 \msun} \right)^{-7/6} \left( \frac{R_{\rm d}}{\amin} \right)^{-7/2} ,
\end{equation}
where the numerical value adopts a radius $R_{\rm d} = \amin$ as motivated by the simulations described in Section \ref{sec:hydrodynamics}. This simple estimate \citep{piran2015-circ,hayasaki2016-spin} suggests that the circularizing shocks are radiatively efficient for $\mh \gtrsim 10^6 \msun$. Based on the density distribution directly obtained from their simulation with $\mh = 2.5 \times 10^6 \msun$, \cite{bonnerot2020-realistic} also find that that the gas optical depths are low enough to allow the radiation generated through shocks to promptly escape. These arguments suggest that the emerging luminosity could be similar to the heating rate with $L_{\rm sh} \approx  \dot{E}_{\rm sh}$, which is likely larger than that originating from the self-crossing shock in most cases. This mechanism therefore represents a plausible explanation for the observed optical and UV luminosities, although this origin may be difficult to distinguish from that associated with gas accretion \citep{metzger2016}. As mentioned in Section \ref{sec:hydrodynamics}, this efficient loss of radiation energy from the system implies that the accretion disc may be initially thinner than expected from the numerical works that assume complete adiabaticity. Radiation-hydrodynamics simulations of this process are required to precisely evaluate the emerging electromagnetic signal and the impact of radiative cooling on the gas evolution.

\subsection{Implications for the nascent accretion flow}

\label{sec:nascent}

The outcome of the complex, three-dimensional hydrodynamical evolution described in Section \ref{sec:hydrodynamics} represents the initial conditions from which accretion onto the black hole will proceed. A detailed analysis of accretion processes is provided in the \diskchap{} and we only emphasize here the properties that can be extracted from the gas evolution at play during disc formation. The main reason for doing so is that most simulations of disc accretion in TDEs are, so far, initialized from an already formed disc, mostly for computational reasons that prevent us from self-consistently following the prior disk assembly process. It is therefore not guaranteed that these frequently used initial conditions possess all the properties described here, and this should be kept in mind by the reader when going through the next chapter.

\subsubsection{Slow gas inflow: no $t^{-5/3}$ decay of the mass accretion rate?}
\label{sec:inflow}

As we have already highlighted, the current picture of disc formation has been significantly improved since the early analytical work by \citet{rees1988}. In contrast to the simple assumption of a fast circularization on scales similar to $\rp$, recent numerical simulations find that disc formation can be a slow process with most of the gas settling at distances $R_{\rm d} \gg \rp$ larger than initially thought \citep{shiokawa2015,sadowski2016,bonnerot2020-realistic}. Assuming a standard $\alpha$-disc, the viscous timescale $\tvisc$ at this location is given by the ratio \citep{frank2002}
\begin{equation}
\frac{\tvisc}{\tmin} \approx \frac{1}{\alpha} \left( \frac{R_{\rm d}}{\amin} \right)^{3/2}  \left( \frac{H}{R} \right)^{-2},
\label{tvisc}
\end{equation}
from which we see that $\tvisc/\tmin \approx 10$ if $R_{\rm d} \approx \amin$. Here, the viscous parameter and the aspect ratio have been given reasonable values of $\alpha \approx 0.1$ and $H/R \approx 1$. This implies that accretion through the disc may be slow, such that a large amount of mass is accumulated at large radii with no significant, time-averaged inflow motion even long after the disc has settled into a quasi-steady state. The early model by \cite{rees1988} proposed that the mass accretion rate follows the same decay law (of $t^{-5/3}$) as the fallback rate, based on the assumptions of rapid disc formation and efficient viscous processing of the freshly returning matter. According to the above analysis, these conditions may not be realized, suggesting a different evolution of the inflow rate through the disc.

\subsubsection{Ballistic accretion and advection: solutions to the ``inverse energy crisis''?}
\label{sec:ballistic}

The total energy radiated in the optical, near-UV, and soft X-ray from most observed TDE candidates is almost always $\lesssim 10^{51} \erg$, a value about three orders of magnitude lower than the rest mass energy of a solar-type star. Additionally, the energy loss necessary for complete circularization is $\mstar \Delta \varepsilon_{\rm circ} \approx 10^{52} \erg$ according to equation \eqref{eq:energy-circular}, which also significantly exceeds the observed values. This results in an ``inverse energy crisis'' where less energy is detected than one would have expected \citep{piran2015-circ,stone2016-rates,lu2018}. We focus below on proposed solutions for this puzzle that involve the hydrodynamics of the accretion flow, but also refer to other possibilities for completeness.

The low energetics of observed events can be explained if most of the gas is ballistically accreted onto the black hole, carrying its orbital energy below the event horizon without any dissipation. As mentioned in Section \ref{sec:hydrodynamics}, ballistic accretion has been found in several simulations \citep{shiokawa2015,sadowski2016,bonnerot2020-realistic} of the disc formation process, which find that a significant fraction of the debris gets accreted with moderate eccentricities. This process is caused by angular momentum redistribution happening during the circularizing shocks and could reduce the radiative efficiency of the flow. A more extreme scenario proposed by \citet{svirski2017} involves an highly eccentric disc with $e\approx \emin$ that is a possible outcome of the disc formation process but has not been found in current simulations. In this situation, magnetic stresses are more efficient at transporting angular momentum near apocenter that acts to progressively reduce the pericenter of the debris. These authors propose that this mechanism can keep operating until all the gas enters the event horizon of the black hole, thereby strongly suppressing the radiation produced by either circularization or accretion. However, another effect of reducing the gas pericenter that close to the gravitational radius is to strongly enhance relativistic apsidal precession. As argued by \citet{bonnerot2017-stream}, the net impact could therefore be to enhance interactions between different parts of this disc, promoting dissipation. A related possibility is that significant dissipation actually occurs inside the forming disc but with a subsequent advection of the injected internal energy, as proposed by \cite{Begelman1979} and \cite{Abramowicz1996}.

Although alleviating the tension, it is not clear whether hydrodynamical models completely solve it and another perhaps more natural  solution relies on the fact that the disc emits in the extreme ultraviolet (EUV) part of the spectrum that is not observable \citep{strubbe2009,roth2016,lu2018,dai2018}. This solution has gained recent support observationally through the detection of infrared dust echoes \citep{van_velzen2016-dust,jiang2016-dust}, Bowen resonance lines \citep{Blagorodnova2019,Leloudas2019} from TDE candidates that are both known to be triggered by extreme-ultraviolet radiation, and direct fitting of disk models to observed X-ray spectra \citep{wen2020}. Indeed, simple models for the infrared dust echoes observed in the TDE flares PTF-09ge and PTF-09axc infer EUV luminosities $\sim10^{44-45}~{\rm erg~s}^{-1}$ \citep{van_velzen2016-dust}. The total EUV energy release from these flares is uncertain, but in both cases could be $\sim 10^{52}~{\rm erg~s}^{-1}$, which would solve the inverse energy crisis if one assumes disruption of lower main sequence stars with masses $M_\star \lesssim 0.3 M_\odot$. While this could be a promising solution, we caution that the modeling of these dust echoes involves many assumptions on the interstellar dust composition and geometry.

\subsubsection{Initial source of disc effective viscosity}

\label{sec:viscosity}

In most accreting systems, angular momentum transport is driven by magnetic stresses resulting from the magneto-rotational instability (MRI) \citep{balbus1991} that are often parametrized in steady state as an effective viscosity. However, it is not obvious that the early phase of accretion in TDEs is produced by the same physical process. This possibility has been put forward by the magneto-hydrodynamical simulations of \citet{sadowski2016}, who found that even after the MRI has reached saturation, angular momentum transport is still dominated by purely hydrodynamical turbulence, which they attribute to either convection or the perturbation created by the matter originally ejected from the self-crossing shock and continuously joining the newly-formed disc.\footnote{We note, however, that over much of the computational domain of \citet{sadowski2016}, the magnetorotational instability is marginally or under-resolved.} Similarly, the simulation by \cite{bonnerot2020-realistic} finds that the forming disc contains spiral shocks excited by later-arriving matter that continuously strikes its outer edge. Angular momentum is transported outward at these locations, which produces a slow inflow through the mid-plane. \citet{nealon2018} also proposed that the gas accretion at early times could be produced by angular momentum transport associated to the Papaloizou-Pringle instability \citep{papaloizou1984}, whose growth timescale was evaluated to be shorter than that of the MRI owing to the low initial disc magnetic field inherited from the disrupted star. Finally, the development of MRI may differ in eccentric discs forming from debris that fails to fully circularize \citep{chan2018}.

\section{Summary and conclusion}
\label{sec:conclusion}

In this chapter, we have presented the current understanding of stream evolution prior to pericenter return (Section \ref{sec:stream-evolution}) and the formation of an accretion flow around the black hole from the bound part of this gas (Section \ref{sec:disc-formation}).

The first phase is well-understood: different stream elements move on ballistic orbits, while their transverse motion is usually at first set by the gas self-gravity (Section \ref{sec:basic}). Several authors have also investigated the influence of various additional physical effects on the stream evolution (Section \ref{sec:additional}). Gravitational fragmentation inside this gas distribution can lead to the formation of clumps, while the confinement by self-gravity can be overcome by either magnetic pressure or through heating associated with hydrogen recombination. It is also possible that the debris experiences interactions with the surrounding medium that can enhance its mixing with this ambient gas. Nevertheless, this additional physics does not appear to drastically change the basic picture of how the stream evolves before its bound part comes back to pericenter.

Our understanding of the second phase of accretion disc formation is less robust. The initial dissipation (Section \ref{sec:initial}) is in most cases dominated by a self-crossing shock whose characteristics are specified by the combined effect of relativistic apsidal precession, expansion from pericenter due to the nozzle shock, and nodal precession produced by the black hole spin. When apsidal precession dominates, local simulations initialized with thin identical stream components find that the collision results in a quasi-spherical expansion of the gas. However, this evolution may significantly change if the nozzle shock imparts a large width difference to the streams or if nodal precession delays the first intersection and causes the collision to be misaligned. Studying the later hydrodynamics requires global simulations that are very numerically challenging owing to the large dynamic range involved, and have therefore only been carried out so far for simplified initial conditions (Section \ref{sec:numerical}). These works find that the debris experiences additional interactions, causing it to progressively move to more circular orbits until eventually settling into an accretion flow (Section \ref{sec:hydrodynamics}). This process appears to complete on a shorter timescale for more relativistic stellar pericenters, due to an increased strength of the initial self-crossing shock. If the gas evolves adiabatically, the final outcome is a thick and extended distribution that typically retains significant internal eccentricities. However, due to the unrealistic setup used in these numerical studies, the precise hydrodynamics at play during accretion flow formation is not yet established and it therefore cannot be excluded that its outcome is different from what current simulations predict. For example,  encounters with weak apsidal precession may lead to a gas distribution that completely fails to circularize and remain instead highly eccentric. As the disc assembles, the radiation produced during circularizing shocks may leave the system to participate to the emerging luminosity from TDEs, especially in the optical/UV bands (Section \ref{sec:shock-driven}). The outcome of the disc formation process represents the initial state for the subsequent accretion onto the black hole and the above works suggest that this later phase consists in a slow inflow of gas induced by effective viscosity with a possibly important contribution from ballistic accretion (Section \ref{sec:nascent}).

Building on the progress made so far, improvements in our understanding of the complex process of accretion flow formation will come in the future from a combination of systematic studies of the different mechanisms involved and global simulations of the entire hydrodynamics applicable to an astrophysically realistic situation. Additional physics missing from most current works such as radiative diffusion, black hole spin and magnetic fields should also be incorporated in these investigations since their influence can significantly affect the gas evolution. Some of these advancements are already being undertaken and will lead to major progress in the theoretical understanding of TDEs.

\begin{acknowledgements}
We gratefully acknowledge conversations with, and detailed comments from T. Piran, as well as his edits on an earlier version of this manuscript.  We are also grateful to E.M. Rossi for insightful discussions during the writing of this chapter.  The research of CB was funded by the Gordon and Betty Moore Foundation through Grant GBMF5076.  NCS was supported by the NASA Astrophysics Theory Research program (grant NNX17AK43G; PI B. Metzger), and from the Israel Science Foundation (Individual Research Grant 2565/19).
\end{acknowledgements}

\bibliography{biblio}
\bibliographystyle{aps-nameyear}          

\end{document}